\documentclass[]{aa}
\usepackage{natbib}         
\usepackage{graphicx}
\usepackage{longtable}
\usepackage[switch]{lineno} 
\usepackage{color}
\usepackage[flushleft]{threeparttable}
\bibpunct{(}{)}{;}{a}{}{,} 

\newcommand{\ind}[1]{_{\mathrm{#1}}}
\newcommand{\diff}{\mathrm{d}}

\newcommand\Dnu{\Delta\nu}

\newcommand\Dnup{\Delta\nu\ind{p}}
\newcommand\DPi{\Delta\Pi\ind{1}}
\newcommand\dnurot{\delta\nu\ind{rot}}

\newcommand\dnurotcore{\delta\nu\ind{rot,core}}

\newcommand\nup{\nu\ind{p}}
\newcommand\nug{\nu\ind{g}}
\newcommand\dl{d_{0\ell}}

\newcommand\numax{\nu\ind{max}}
\newcommand\nmax{n\ind{max}}

\newcommand\DP{\Delta P}

\newcommand\Dtaum{\Delta\tau\ind{m}}

\newcommand\np{{n\ind{p}}}
\newcommand\ngrav{{n\ind{g}}}

\newcommand\epsp{\varepsilon\ind{p}}

\newcommand\N{\mathcal{N}}

\newcommand{\BV}{Brunt-V\"ais\"al\"a \,}
\newcommand{\NBV}{N\ind{BV}}

\newcommand\Teff{T\ind{eff}}
\newcommand\Msol{{M\ind{\odot}}}
\newcommand\Rsol{{R\ind{\odot}}}
\newcommand\Tsol{{T\ind{\odot}}}

\newcommand\numaxsol{{\nu\ind{max,\odot}}}
\newcommand\Dnusol{\Dnu\ind{\odot}}

\newcommand\bcore{\beta\ind{core}}
\newcommand\benv{\beta\ind{env}}
\newcommand\omegacore{\langle\Omega\rangle\ind{core}}
\newcommand\omegaenv{\langle\Omega\rangle\ind{env}}

\begin{document}

\author{%
 C. Gehan\inst{1},
 B. Mosser\inst{1},
 E. Michel\inst{1},
R. Samadi\inst{1},
T. Kallinger\inst{2}
 }

\institute{
 PSL, LESIA, CNRS, Universit\'e Pierre et Marie Curie,
 Universit\'e Denis Diderot, Observatoire de Paris, 92195 Meudon
 cedex, France; \texttt{charlotte.gehan@obspm.fr}
\and Institute for Astrophysics, University of Vienna, Türkenschanzstrasse 17, 1180 Vienna, Austria
 }


\abstract{Asteroseismology allows us to probe stellar interiors. In the case of red giant stars, conditions in the stellar interior are such to allow for the existence of mixed modes, consisting in a coupling between gravity waves in the radiative interior and pressure waves in the convective envelope. Mixed modes can thus be used to probe the physical conditions in red giant cores. However, we still need to identify the physical mechanisms that transport angular momentum inside red giants, leading to the slow-down observed for the red giant core rotation. Thus large-scale measurements of the red giant core rotation are of prime importance to obtain tighter constraints on the efficiency of the internal angular momentum transport, and to study how this efficiency changes with
stellar parameters.}
{This work aims at identifying the components of the rotational multiplets for dipole mixed modes in a large number of red giant oscillation spectra observed by \textit{Kepler}. Such identification provides us with a direct measurement of the red giant mean core rotation.}
{We compute stretched spectra that mimic the regular pattern of pure dipole gravity modes. Mixed modes with same azimuthal order are expected to be almost equally spaced in stretched period, with a spacing equals to the pure dipole gravity mode period spacing. The departure from this regular pattern allows us to disentangle the various rotational components and therefore to determine the mean core rotation rates of red giants.}
{We automatically identify the rotational multiplet components of 1183 stars on the red giant branch with a success rate of 69\% with respect to our initial sample. As no information on the internal rotation can be deduced for stars seen pole-on, we obtain mean core rotation measurements for 875 red giant branch stars. This large sample includes stars with a mass as large as 2.5 $\Msol$, allowing us to test the dependence of the core slow-down rate on the stellar mass.}
{Disentangling rotational splittings from mixed modes is now possible in an automated way for stars on the red giant branch, even for the most complicated cases, where the rotational splittings exceed half the mixed-mode period spacing. This work on a large sample allows us to refine previous measurements of the evolution of the mean core rotation on the red giant branch. Rather than a slight slow down, our results suggest rotation to be constant along the red giant branch, with values independent on the mass.}


\title{Core rotation braking on the red giant branch \newline for various mass ranges}
\authorrunning{C. Gehan et al.}
\titlerunning{Core rotation braking on the red giant branch for various mass ranges}
\maketitle

\section{Introduction}

The ultra-high precision photometric space missions CoRoT and \textit{Kepler} have recorded extremely long observation runs, providing us with seismic data of unprecedented quality. The surprise came from red giant stars (e.g., \cite{Mosser_2016}), presenting solar-like oscillations which are stochastically excited in the external convective envelope \citep{Dupret}. Oscillation power spectra showed that red giants not only present pressure modes as in the Sun, but also mixed modes \citep{De_Ridder, Bedding} resulting from a coupling of pressure waves in the outer envelope with internal gravity waves \citep{Scuflaire}. As mixed modes behave as pressure modes in the convective envelope and as gravity modes in the radiative interior, they allow one to probe the core of red giants \citep{Beck}.
\newline
Dipole mixed modes are particularly interesting because they are mostly sensitive to the red giant core \citep{Goupil}. They were used to automatically measure the dipole gravity mode period spacing $\DPi$ for almost 5000 red giants \citep{Vrard}, providing information about the size of the radiative core \citep{Montalban} and defined as
\begin{equation}\label{DPi1}
\DPi = \frac{2 \pi^2}{\sqrt{2}} \left( \int_{core} \frac{\NBV}{r} \diff r \right)^{-1},
\end{equation}
where $\NBV$ is the \BV frequency.
The measurement of $\DPi$ leads to the accurate determination of the stellar evolutionary stage and allows us to distinguish shell-hydrogen burning red giants from core-helium burning red giants \citep{Bedding, Stello, Mosser_2014}.
\newline
Dipole mixed modes also give access to near-core rotation rates \citep{Beck}. Rotation has been shown to impact not only the stellar structure by perturbing the hydrostatic equilibrium, but also the internal dynamics of stars by means of the transport of both angular momentum and chemical species \citep{Zahn, Talon, Lagarde}. It is thus crucial to measure this parameter for a large number of stars to monitor its effect on stellar evolution \citep{Lagarde_2016}. Semi-automatic measurements of the mean core rotation of about 300 red giants indicated that their cores are slowing down along the red giant branch while contracting at the same time \citep{Mosser_2012c, Deheuvels}. Thus, angular momentum is efficiently extracted from red giant cores \citep{Eggenberger, Cantiello}, but the physical mechanisms supporting this angular momentum transport are not yet fully understood. Indeed, several physical mechanisms transporting angular momentum have been implemented in stellar evolutionary codes, such as meridional circulation and shear turbulence \citep{Eggenberger, Marques, Ceillier}, mixed modes \citep{Belkacem_2015a, Belkacem_2015b}, internal gravity waves \citep{Fuller, Pincon}, and magnetic fields \citep{Cantiello, Rudiger}, but none of them can reproduce the measured orders of magnitude for the core rotation along the red giant branch. In parallel, several studies have tried to parameterize the efficiency of the angular momentum transport inside red giants through ad-hoc diffusion coefficients \citep{Spada, Eggenberger_2017}.
\newline
In this context, we need to obtain mean core rotation measurements for a much larger set of red giants in order to put stronger constraints on the efficiency of the angular momentum transport 
and to study how this efficiency changes with the global stellar parameters like mass \citep{Eggenberger_2017}. In particular, we require measurements for stars on the red giant branch because the dataset anyalysed by \cite{Mosser_2012c} only includes 85 red giant branch stars. The absence of large-scale measurements is due to the fact that rotational splittings often exceed half the mixed-mode frequency spacings at low frequencies in red giants. For such conditions, disentangling rotational splittings from mixed modes is challenging. Nevertheless, it is of prime importance to develop a method as automated as possible as we entered the era of massive photometric data, with \textit{Kepler} providing light curves for more than 15 000 red giant stars and the future Plato mission potentially increasing this number to hundreds of thousands.
\newline
In this work we set up an almost fully automated method to identify the rotational signature of stars on the red giant branch. Our method is not suitable to clump stars, presenting smaller rotational splittings as well as larger mode widths due to shorter lifetimes of gravity modes \citep{Vrard_2017}. Thus, the analysis of the core rotation of clump stars is beyond the scope of this paper. In Section \ref{principle} we explain the principle of the method, based on the stretching of frequency spectra to obtain period spectra reproducing the evenly-spaced gravity-mode pattern. In Section \ref{method} we detail the set up of the method, including the estimation of the uncertainties. In Section \ref{comparison} we compare our results with those obtained by \cite{Mosser_2012c}. In Section \ref{results} we apply the method to red giant branch stars of the \textit{Kepler} public catalog, and investigate the impact of the stellar mass in the evolution of the core rotation. Section \ref{discussion} is devoted to discussion and Section \ref{conclusion} to conclusions.

\section{Principle of the method}\label{principle}

Mixed modes have a dual nature: pressure-dominated mixed-modes (p-m modes) are almost equally spaced in frequency, with a frequency spacing close to the large separation $\Dnu$, while gravity-dominated mixed-modes (g-m modes) are almost equally spaced in period, with a period spacing close to $\DPi$.
In order to retrieve the behavior of pure gravity modes, we need to disentangle the different contributions of p-m and g-m modes, which can be done by deforming the frequency spectra.

\subsection{Stretching frequency spectra}

\begin{figure}
\resizebox{\hsize}{!}{\includegraphics{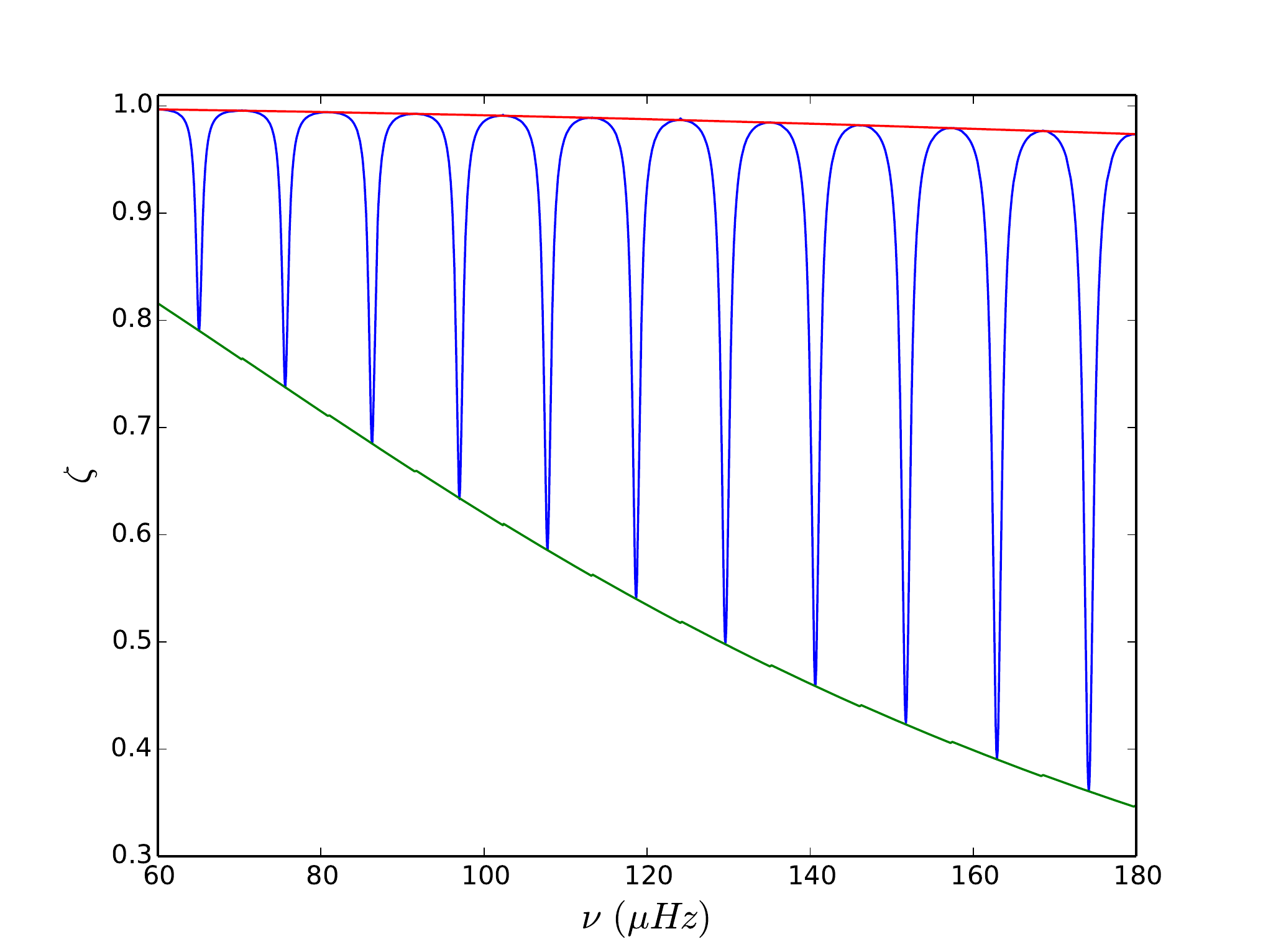}}
\caption{The stretching function $\zeta$ (blue line) computed with $\Dnu = 11$ $\mu$Hz, $\DPi=80$ s, and $q=0.12$. The asymptotic behaviors of $\zeta$ for g-m and p-m modes are indicated by a red and green line, respectively.}
\label{fig-zeta}
\end{figure}

The mode frequencies of pure pressure modes are estimated through the red giant universal oscillation pattern \citep{Mosser_2011}:
\begin{equation}\label{eqt-p-2eordre}
\nu\ind{p,\ell} = \left(\np + \frac{\ell}{2} + \epsp + \dl + \frac{\alpha}{2} (\np - \nmax)^2 \right) \Dnu,
\end{equation}
where:
\begin{itemize}
\item $\np$ is the pressure radial order;
\item $\ell$ is the angular degree of the oscillation mode;
\item $\epsp$ is the phase shift of pure pressure modes;
\item $\alpha$ represents the curvature of the radial oscillation pattern;
\item $\dl$ is the small separation, namely the distance, in units of $\Dnu$, of the pure pressure mode having an angular degree equal to $\ell$, compared to the midpoint between the surrounding radial modes;
\item $\nmax = \numax / \Dnu - \epsp$ is the non-integer order at the frequency $\numax$ of maximum oscillation signal.
\end{itemize}
We consider only dipole mixed modes, which are mainly sensitive to the red giant core. Thus, we remove in a first step the frequency ranges where radial and quadrupole modes are expected from the observed spectra using the universal oscillation pattern (Eq.~\ref{eqt-p-2eordre}).
\newline
 We then convert the frequency spectra containing only dipole mixed modes into stretched period spectra, with the stretched period $\tau$ derived from the differential equation \citep{Mosser_2015}
\begin{equation}\label{eqt-asymp}
	\diff \tau = \frac{1}{\zeta} \frac{\diff \nu}{\nu^2}.
\end{equation}
The $\zeta$ function is defined as (Margarida Cunha, private communication)
\begin{equation}\label{eqt-zeta}
    \zeta = \left[1 + \frac{\nu^2}{q} \frac{\DPi}{\Dnup} \frac{1}{\frac{1}{q^2} \sin^2 \left(\pi \frac{\nu - \nup}{\Dnup}\right) + \cos^2 \left(\pi \frac{\nu - \nup}{\Dnup}\right)}\right]^{-1}.
\end{equation}
The parameters entering the definition of $\zeta$ are:
\begin{itemize}
\item $q$ the coupling parameter between gravity and pressure modes;
\item $\Dnup = \Dnu \left( 1 + \alpha (\np - \nmax) \right)$ the observed large separation which increases with the radial order;
\item $\nup$ the pure dipole pressure mode frequencies;
\item $\nu$ the mixed-mode frequencies.
\end{itemize}
The pure dipole pressure mode frequencies are given by Eq.~\ref{eqt-p-2eordre} for $\ell = 1$.
The mixed-mode frequencies are given by the asymptotic expansion of mixed modes \citep{Mosser_2012a}
\begin{equation}\label{eqt-mixed-modes}
\nu = \nup + \frac{\Dnup}{\pi} \arctan \left[q \tan \pi \left( \frac{1}{\DPi \nu} - \frac{1}{\DPi \nug} \right) \right],
\end{equation}
where
\begin{equation}\label{Dnu-np}
\nug = \frac{1}{\ngrav \DPi}
\end{equation}
are the pure dipole gravity mode frequencies, with $\ngrav$ being the gravity radial order, usually defined as a negative integer.
\newline
\newline
We can approximate
\begin{equation}\label{zeta-DP-DPi}
\zeta \simeq \frac{\DP}{\DPi},
\end{equation}
where $\DP$ is the bumped period spacing between consecutive dipole mixed modes. In these conditions, $\zeta$ gives information on the trapping of dipole mixed modes. Gravity-dominated mixed-modes (g-m modes) have a period spacing close to $\DPi$ and represent the local maxima of $\zeta$, while pressure-dominated mixed-modes (p-m modes) have a lower period spacing decreasing with frequency and represent the local minima of $\zeta$ (Fig.~\ref{fig-zeta}).

\begin{figure*}
\centering
\includegraphics[width=6cm]{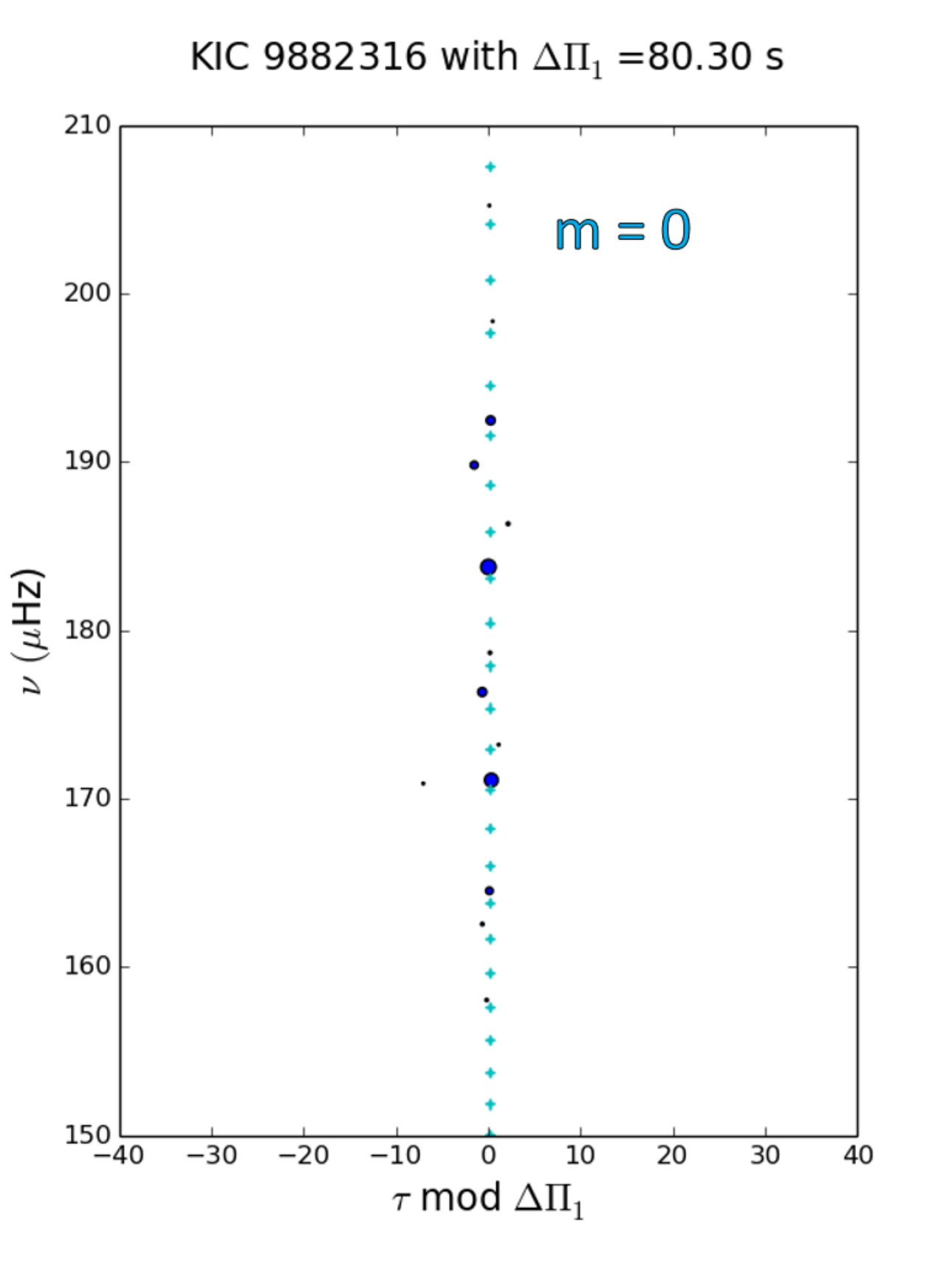}
\includegraphics[width=6cm]{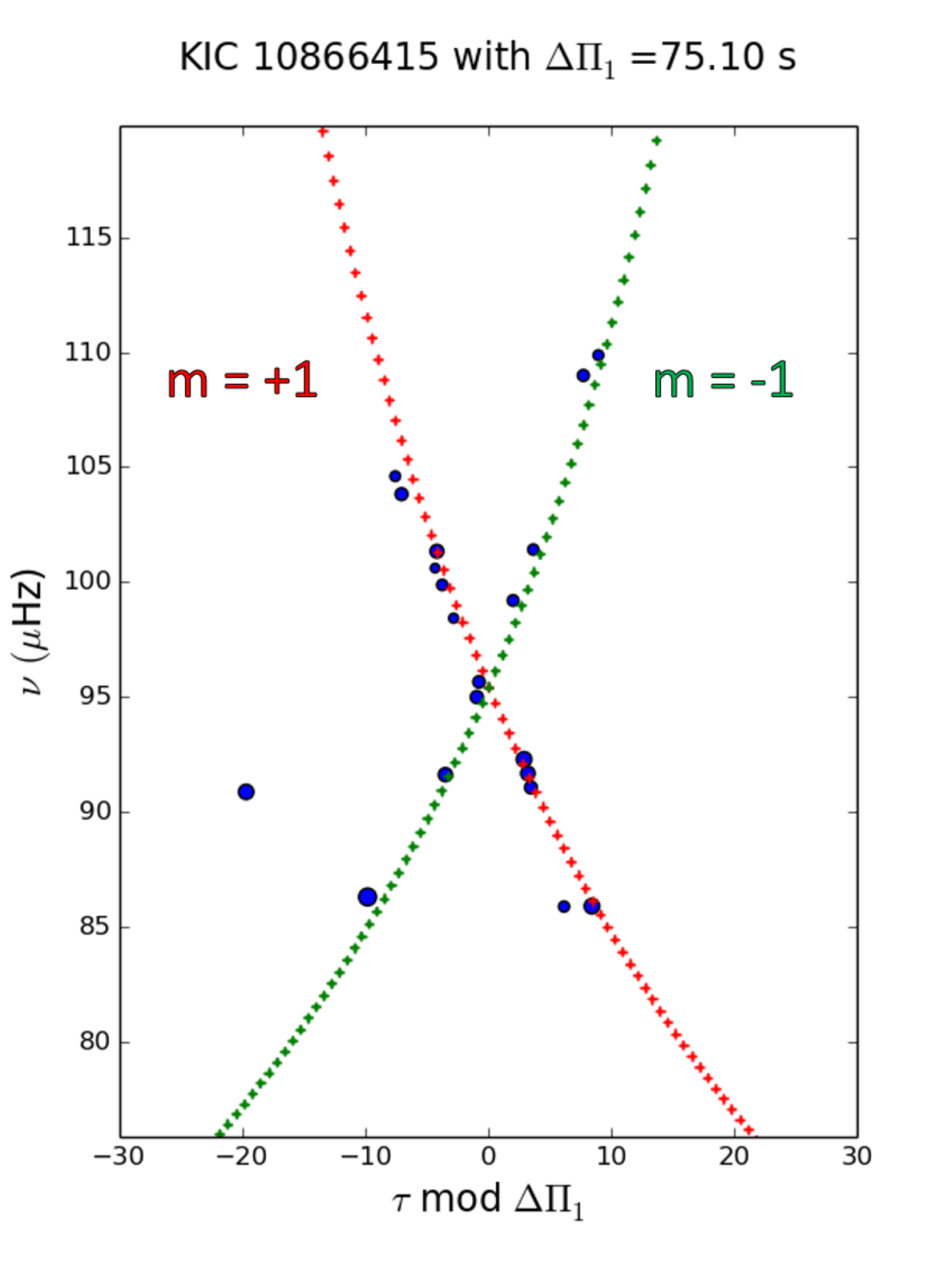}
\includegraphics[width=6cm]{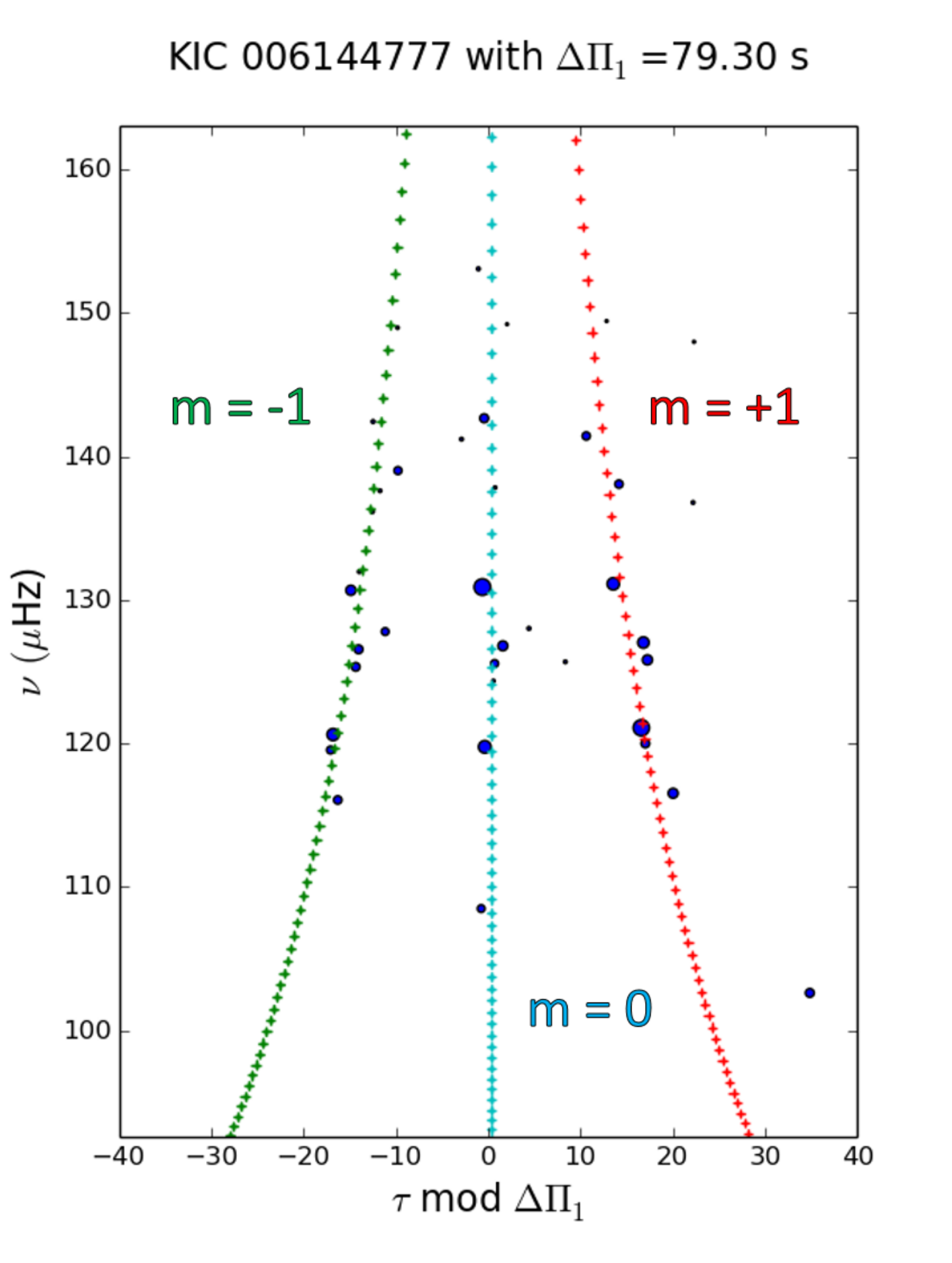}
\caption{Stretched period échelle diagrams for red giant branch stars with different inclinations. The rotational components are identified in an automatic way through a correlation of the observed spectrum with a synthetic one constructed using Eq.~\ref{eqt-spacing-rot}. The colors indicate the azimuthal order: the $m=\lbrace-1,0,+1\rbrace$ rotational components are represented in green, light blue and red, respectively. The symbol size of the observations increases with the power spectral density. \textit{Left:} Star seen pole-on where only the $m=0$ rotational component is visible. \textit{Middle:} Star seen equator-on where the $m=\{-1,+1\}$ components are visible. \textit{Right:} Star having an intermediate inclination angle where all the three components are visible.}
\label{fig-echelle}
\end{figure*}

\subsection{Revealing the rotational components}

P-m modes are not only sensitive to the core but also to the envelope. The next step thus consists in removing p-m modes from the spectra through Eq.~\ref{eqt-p-2eordre}, leaving only g-m modes. 
In practice, g-m modes with a  height-to-background ratio larger than 6 are considered as significant in a first step; then this threshold is manually adapted in order to obtain the best compromise between the number of significant modes and background residuals. 
All dipole g-m modes with same azimuthal order $m$ should then be equally spaced in stretched period, with a spacing close to $\DPi$. However, the core rotation perturbs this regular pattern so that the stretched period spacing between consecutive g-m modes with same azimuthal order slightly differs from $\DPi$, with the small departure depending on the mean core rotational splitting as \citep{Mosser_2015}
\begin{equation}\label{eqt-spacing-rot-ini}
 \Dtaum = \DPi \left( 1 + 2 \, m \, \zeta \, \frac{\dnurotcore}{\nu} \right).
\end{equation}
As the mean value of $\zeta$ depends on the mixed-mode density $\N$, we can avoid the calculation of $\zeta$ by approximating
\begin{equation}\label{eqt-spacing-rot}
\Dtaum \simeq \DPi \left( 1 + 2 \, m \, \frac{\N}{\N+1} \, \frac{\dnurotcore}{\nu} \right).
\end{equation}
The mixed-mode density represents the number of gravity modes per $\Dnu$-wide frequency range and is defined as
\begin{equation}\label{eqt-nmix}
    \N = {\Dnu \over \DPi \, \numax^2}.
\end{equation}
\newline
Red giants are slow rotators, presenting low rotation frequencies of the order of 2 $\mu$Hz or less. In these conditions, $\Omega(r)/2 \pi < \Dnu \ll \numax$ throughout the whole star and rotation can be treated as a first-order perturbation of the hydrostatic equilibrium \citep{Ouazzani}. Thus, the rotational splitting can be written as (\cite{Unno}, see also \cite{Goupil} for the case of red giants)
\begin{equation}\label{dnurot-1}
\dnurot = \int_{0}^{1} K(x) \frac{\Omega(x)}{2 \pi} \diff x,
\end{equation}
where $x = r / R$ is the normalized radius, $K(x)$ is the rotational kernel and $\Omega(x)$ is the angular velocity at normalized radius $x$.
In first approximation we can separate the core and envelope contributions of the rotational splitting as
\begin{equation}\label{dnurot-2}
\dnurot =\bcore \left\langle \frac{\Omega}{2 \pi} \right\rangle \ind{core} + \benv \left\langle \frac{\Omega}{2 \pi} \right\rangle \ind{env},
\end{equation}
with
\begin{equation}\label{bcore}
\bcore = \int_{0}^{x \ind{core}} K(x) \diff x,
\end{equation}
\begin{equation}\label{benv}
\benv = \int_{x \ind{core}}^{1} K(x) \diff x,
\end{equation}
\begin{equation}\label{omegacore}
\omegacore = \frac{\int_{0}^{x\ind{core}} \Omega(x) K(x) \diff x}{\int_{0}^{x\ind{core}} K(x) \diff x},
\end{equation}
\begin{equation}\label{omegaenv}
\omegaenv = \frac{\int_{x\ind{core}}^{1} \Omega(x) K(x) \diff x}{\int_{x\ind{core}}^{1} K(x) \diff x},
\end{equation}
$x \ind{core} = r \ind{core} / {R}$ being the normalized radius of the g-mode cavity.
\newline
If we assume that dipole g-m modes are mainly sensitive to the core, we can neglect the envelope contribution as $\bcore / \benv \gg 1$ in these conditions. Moreover, $\bcore \simeq 1/2$ for dipole g-m modes \citep{Ledoux}.
In these conditions, a linear relation connects the core rotational splitting to the mean core angular velocity as \citep{Goupil, Mosser_2015}
\begin{equation}\label{omega}
\dnurotcore \simeq \frac{1}{2} \left\langle \frac{\Omega}{2 \pi} \right\rangle \ind{core}.
\end{equation}
The departure of $\Dtaum$ from $\DPi$ is small, on the order of a few percent of $\DPi$. This allows us to fold the stretched spectrum with $\DPi$ in order to build stretched period échelle diagrams \citep{Mosser_2015}, in which the individual rotational components align according to their azimuthal order and become easy to identify (Fig.~\ref{fig-echelle}). When the star is seen pole-on, the $m = 0$ components line up on an unique and almost vertical ridge. Rotation modifies this scheme by splitting mixed modes into two or three components, depending on the stellar inclination.
\newline
In these échelle diagrams, rotational splittings and mixed modes are now disentangled. It is possible to identify the azimuthal order of each component of a rotational multiplet, even in a complex case as KIC 10866415 where the rotational splitting is much larger than the mixed-mode frequency spacing. In such cases, the ridges cross each other (Fig.~\ref{fig-echelle}).


\section{Disentangling and measuring rotational splittings}\label{method}

\begin{figure}
\centering
\resizebox{\hsize}{!}{\includegraphics{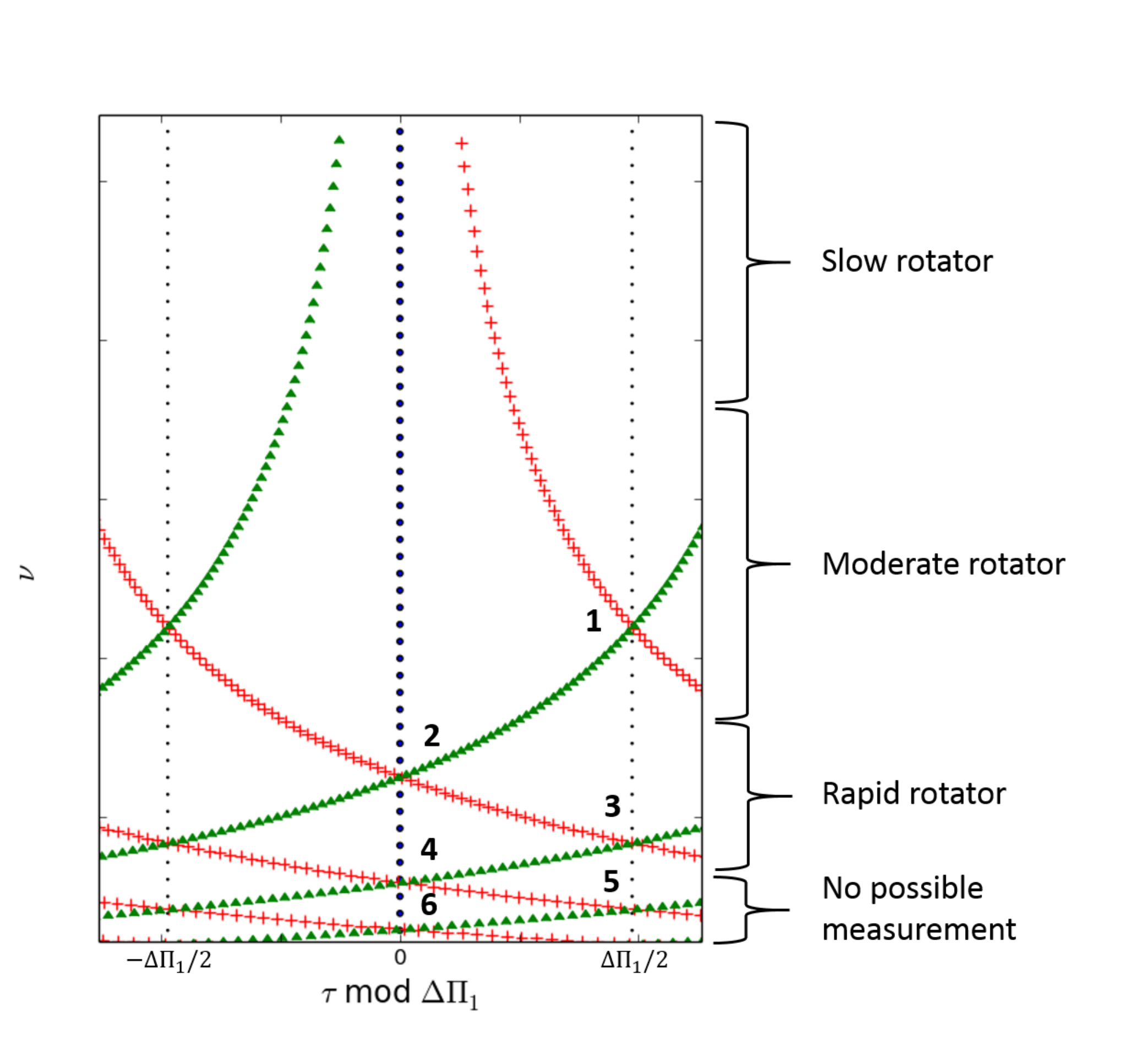}}
\caption{Synthetic stretched period échelle diagram built from Eqs.~(\ref{eqt-spacing-rot},\ref{eqt-nmix}). The colors indicate the azimuthal order: the $m=\lbrace-1,0,+1\rbrace$ rotational components are given by green triangles, blue dots and red crosses, respectively. The small black dots represent $\tau=\left(\pm\DPi/2\right)$ mod $\DPi$. The numbers mark the crossing order $k$ (Eq.~\ref{eqt-crossing}). As the number of observable modes is limited, they cover only a limited frequency range in the diagram.}
\label{fig-synthetic}
\end{figure}

We developed a method allowing for an automated identification of the rotational multiplet components with stretched period échelle diagrams. The method is based on a correlation between the observed spectrum and a synthetic one built from Eqs.~(\ref{eqt-spacing-rot},\ref{eqt-nmix}) \citep{Gehan_2017}. Fig.~\ref{fig-synthetic} shows an example of such a synthetic échelle diagram, where the rotational components present many crossings. These crossings between multiplet rotational components with different azimuthal orders are similar to what was shown by \cite{Ouazzani} for theoretical red giant spectra. However, they occur here at a smaller core pulsation frequency, for $200 < \langle\Omega\ind{core}\rangle / 2 \pi < 2000$ nHz, while \cite{Ouazzani} found the first crossings to happen around $\langle\Omega\ind{core}\rangle / 2 \pi = 8 \, \mu$Hz.
\newline
\newline
The frequencies where crossings occur can be expressed as \citep{Gehan_2017}
\begin{equation}\label{eqt-crossing}
    \nu\ind{k} = \sqrt{\frac{\dnurotcore}{k \DPi}},
\end{equation}
where $k$ is a positive integer representing the crossing order.
The échelle diagram shown in Fig.~\ref{fig-synthetic} can be understood as follows. The upper part corresponds to slow rotators with no crossing, which can be found on the lower giant branch. The medium part, where the first crossing occurs, corresponds to moderate rotators. Such complicated cases are found at lower frequencies, corresponding to most of the evolved stars in the \textit{Kepler} sample, where rotational components can now be clearly disentangled through the use of the stretched period. The lower part corresponds to rapid rotators, where too many crossings occur to allow the identification of the multiplet components in the unperturbed frequency spectra. Nevertheless, rotational components can still be disentangled using stretched period spectra. The lowest part corresponds to very rapid rotators with too many crossings to allow the identification of the multiplet rotational components, for which currently no measurement of the core rotation is possible.
\newline
If we consider stars having the same $\DPi$, the larger the rotational splittings, the higher the crossing frequencies, and, in average, the larger the number of visible crossings.
\newline
\newline
Such synthetic échelle diagrams were originally used to identify the crossing order for stars with overlapping multiplet rotational components \citep{Gehan_2017}. Combined with the measurement of at least one frequency where a crossing occurs, the identification of the crossing order provides us with a measurement of the mean core rotational splitting. In this study, we base our work on this method, extending it to allow for the identification of the rotational multiplet components whether these components overlap or not.

\subsection{Correlation of the observed spectrum with synthetic ones}
The observed spectrum is correlated with different synthetic spectra constructed with different $\dnurotcore$ and $\DPi$ values, through an iterative process. The position of the synthetic spectra is based on the position of important peaks in the observed spectra, defined as having a power spectral density greater than or equal to 0.25 times the maximal power spectral density of g-m modes.
\newline 
We test $\dnurotcore$ values ranging from 100 nHz to $1 \, \mu$Hz with steps of 5 nHz. In fact, the method is not adequate for $\dnurotcore<100$ nHz and for $\dnurotcore > 1 \, \mu$Hz: when $\dnurotcore$ is too low, the multiplet rotational components are too close to be unambiguously distinguished in échelle diagrams; when $\dnurotcore$ is too high, the multiplet rotational components overlap over several g-m mode orders, making their identification challenging.
\newline
We also considered $\DPi$ as a flexible parameter. We used the $\DPi$ measurements of \cite{Vrard} as a first guess and tested $\Delta\Pi\ind{1,test}$ in the range $\DPi \, (1 \pm \, 0.03)$ with steps of 0.1 s.
Indeed, inaccurate $\DPi$ measurements can occur when only a low number of g-m modes are observed, due to suppressed dipole modes \citep{Mosser_2017} or high up the giant branch where g-m modes have a very high inertia and cannot be observed \citep{Grosjean}. Moreover, even small variations of the folding period $\DPi$ modify the inclination of the observed ridges in the échelle diagram. They are then no longer symmetric with respect to the vertical $m=0$ ridge if the folding period $\DPi$ is not precise enough, and the correlation with synthetic spectra may fail. Thus, our correlation method does not only give high-precision $\dnurotcore$ measurements, but also allows us to improve the precision on the measurement of $\DPi$, expected to be as good as 0.01 \%.
\newline
Furthermore, the stellar inclination is not known a priori and impacts the number of visible rotationally split frequency components in the spectum. Thus, three types of synthetic spectra are tested at each time step, containing one, two and three rotational multiplet components.

\begin{figure}
\centering
\resizebox{\hsize}{!}{\includegraphics[width=9.1cm]{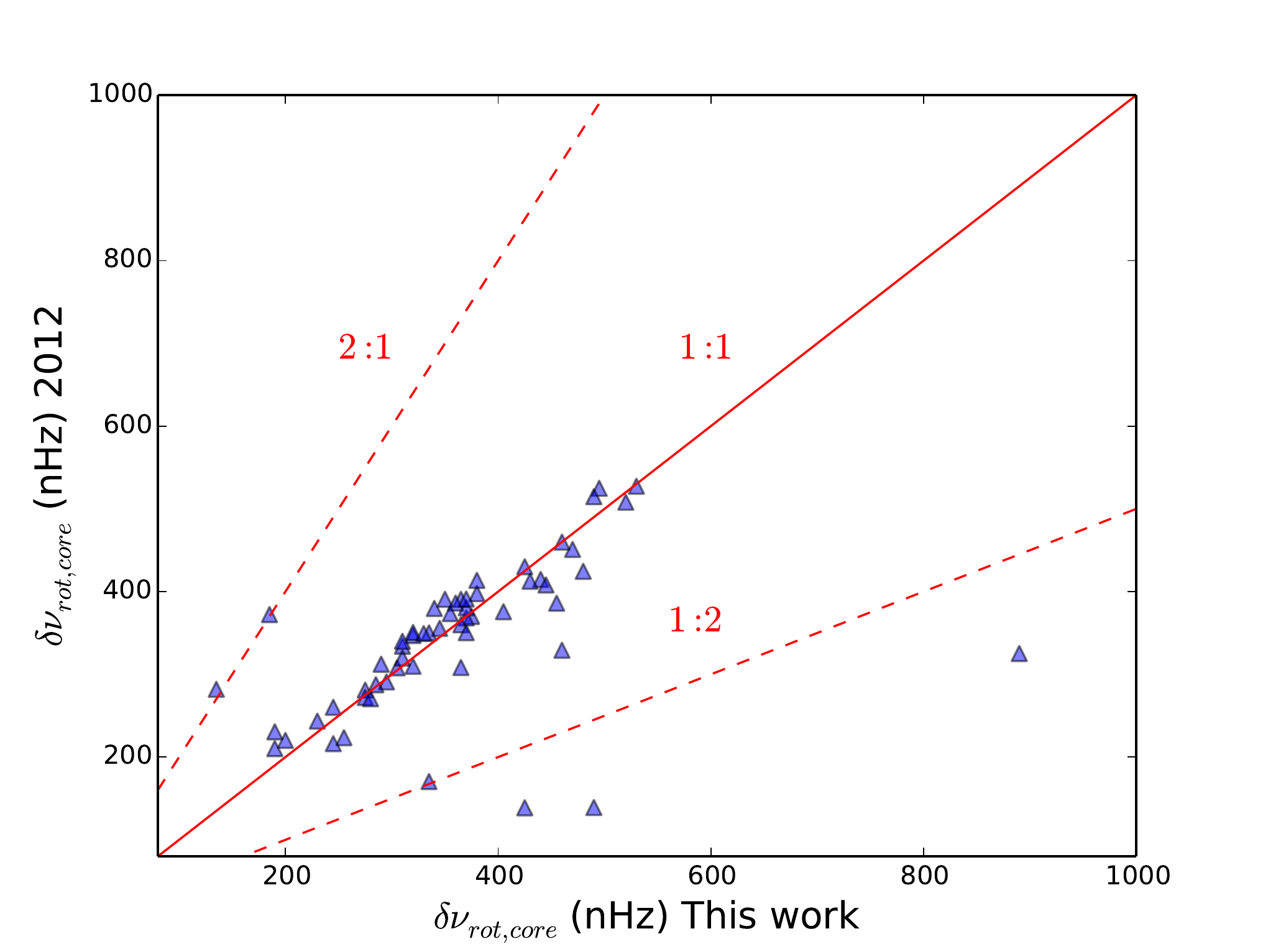}}
\caption{Comparison of the present results with those from \cite{Mosser_2012c} for stars on the red giant branch. The solid red line represents the 1:1 relation. The red dashed-lines represent the 1:2 and 2:1 relations.}
\label{fig-comp-2012}
\end{figure}

\begin{figure*}
\centering
\includegraphics[width=12cm]{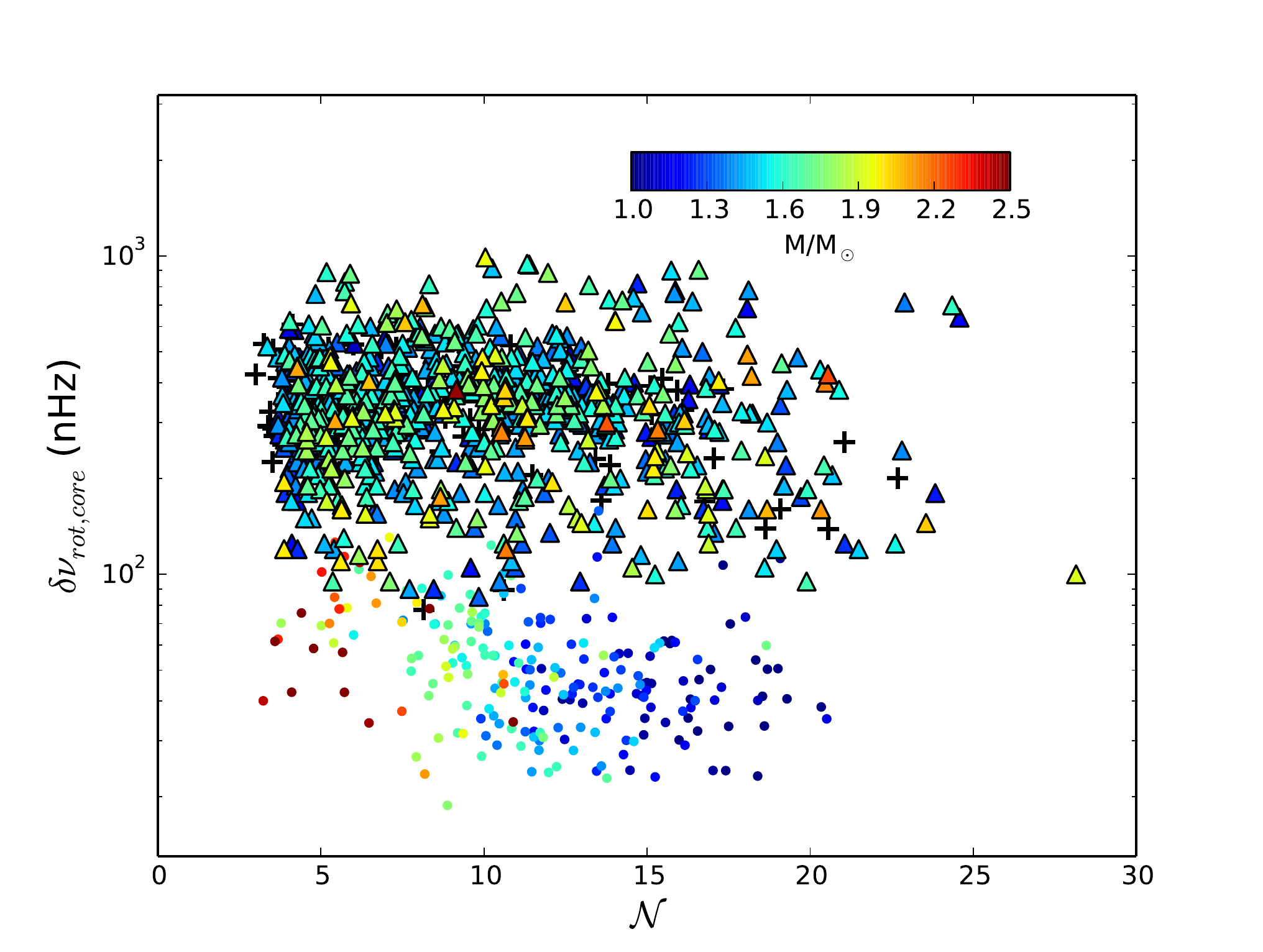}
\caption{Mean core rotational splitting as a function of the mixed-mode density measured through Eq.~\ref{eqt-nmix}. The color code indicates the stellar mass estimated from the asteroseismic global parameters. Colored triangles represent the measurements obtained in this study. Black crosses and colored dots represent the measurements of \cite{Mosser_2012c} for stars on the red giant branch and in the red clump, respectively.}
\label{fig-dnurot}
\end{figure*}

\subsection{Selection of the best-fitting synthetic spectrum}

For each configuration tested, namely for synthetic spectra including one, two or three components, the best fit is selected in an automated way by maximizing the number of peaks aligned with the synthetic ridges. After this automated step, an individual check is performed to select the best solution, depending on the number of rotational components. This manual operation allows us to correct for the spurious signatures introduced by short-lived modes or $\ell=3$ modes.
\newline
\newline
We define $\tau\ind{peak}$ and $\tau\ind{synt}$ the observed and synthetic stretched periods of any given peak, respectively. We empirically consider that a peak is aligned with a synthetic ridge if
\begin{equation}\label{eqt-peaks-alignment}
\left| \tau\ind{peak} - \tau\ind{synth} \right| \leq \frac{\DPi}{30}.
\end{equation}
We further define
\begin{equation}\label{eqt-R2}
\chi^{2} = \frac{\sum\ind{i=1}^{n}\left({\tau\ind{peak,i} - \tau\ind{synth,i}}\right)^2}{n},
\end{equation}
where $n$ is the total number of aligned peaks along the synthetic ridges. $\chi^{2}$ is the mean residual squared sum for peaks belonging to the synthetic rotational multiplet components. This quantity thus represents an estimate of the spread of the observed $\tau\ind{peak}$ values around the synthetic $\tau\ind{synth}$. If several $\dnurotcore$ and $\Delta\Pi\ind{1,test}$ values give the same number of aligned peaks, then the best fit corresponds to the minimimum $\chi^{2}$ value.
\newline
This step provides us with the best values of $\Delta\Pi\ind{1,test}$ and $\dnurotcore$ for each of the three stellar inclinations tested. At this stage at most three possible synthetic spectra remain, corresponding to fixed $\Delta\Pi\ind{1,test}$ and $\dnurotcore$ values: a spectrum with one, two or three rotational multiplet components.
The final fit corresponding to the actual configuration is selected manually. Such final fits shown in Fig.~\ref{fig-echelle} provide us with a measurement of the mean core rotational splitting, except when the star is nearby pole-on.

\subsection{Uncertainties}

The uncertainty $\sigma$ on the measurement of $\dnurotcore$ is calculated through Eq.~\ref{eqt-R2}, except that peaks with 
\begin{equation}\label{eqt-peaks-signifiance}
\left| \tau\ind{peak} - \tau\ind{synth} \right| > \frac{\DPi}{3}
\end{equation}
are considered as non-significant and are discarded from the estimation of the uncertainties. They might be due to background residuals, to $\ell=1$ p-m modes that were not fully discarded or to $\ell=3$ modes.
The obtained uncertainties are on the order of $10$ nHz or smaller.

\section{Comparison with other measurements}\label{comparison}

\cite{Mosser_2012c} measured the mean core rotation of about 300 stars, both on the red giant branch and in the red clump. We apply our method to the red giant branch stars of \cite{Mosser_2012c} sample, representing 85 stars. The method proposed a satisfactory identification of the rotational components in 79$\%$ of cases. Nevertheless, for some cases our method detects only the $m = 0$ component while \citet{Mosser_2012c} obtained a measurement of the mean core rotation, which requires the presence of $m = \pm1$ components in the observations. Finally, our method provides $\dnurotcore$ measurements for 67\% of the stars in the sample. Taking as a reference the measurements of \cite{Mosser_2012c}, we can thus estimate that we miss about  12$\%$ of the possible $\dnurotcore$ measurements by detecting only the $m=0$ rotational component while the $m=\pm \, 1$ components are also present but remain undetected by our method. This mostly happens at low inclination values, when the visibility of the $m=\pm \, 1$ components is very low compared to that of $m=0$. In such cases, the components associated to $m = \pm \, 1$ appear to us lost in the background noise.
\newline
We calculated the relative difference between the present measurements and those of \cite{Mosser_2012c} as
\begin{equation}\label{eqt-rel-diff}
 d\ind{r} = \left| \frac{\dnurot - \delta\nu\ind{rot,2012}}{\delta\nu\ind{rot,2012}} \right|.
\end{equation}
 We find that $d\ind{r} < 10\%$ for 74$\%$ of stars (Fig.~\ref{fig-comp-2012}). We expect our correlation method to provide more accurate measurements because we use stretched periods based on $\zeta$, while \cite{Mosser_2012c} chose a Lorentzian profile to reproduce the observed modulation of rotational splittings with frequency, which has no theoretical basis.
\newline
 On the 57 $\dnurotcore$ measurements that we obtained for the RGB stars in the \citet{Mosser_2012c} sample, only 7 lie away from the 1:1 comparison. We can easily explain this discrepancy for 3 of these stars, lying either on the lines representing a 1:2 or a 2:1 relation. Indeed, in some cases the rotational multiplet components are entangled at low frequency when the rotational splitting exceeds half the mixed-mode period spacing. It is thus easy to misidentify the components and measure either half or twice the rotational splitting. Our method based on stretched periods deals with complicated cases corresponding to large splittings with more accuracy compared to \citet{Mosser_2012c}, avoiding a misidentification of the rotational components. We note that \citet{Mosser_2012c} measured maximum splittings values around $600$ nHz while our measurements indicate values as high as $900$ nHz.

\begin{figure}
\resizebox{\hsize}{!}{\includegraphics{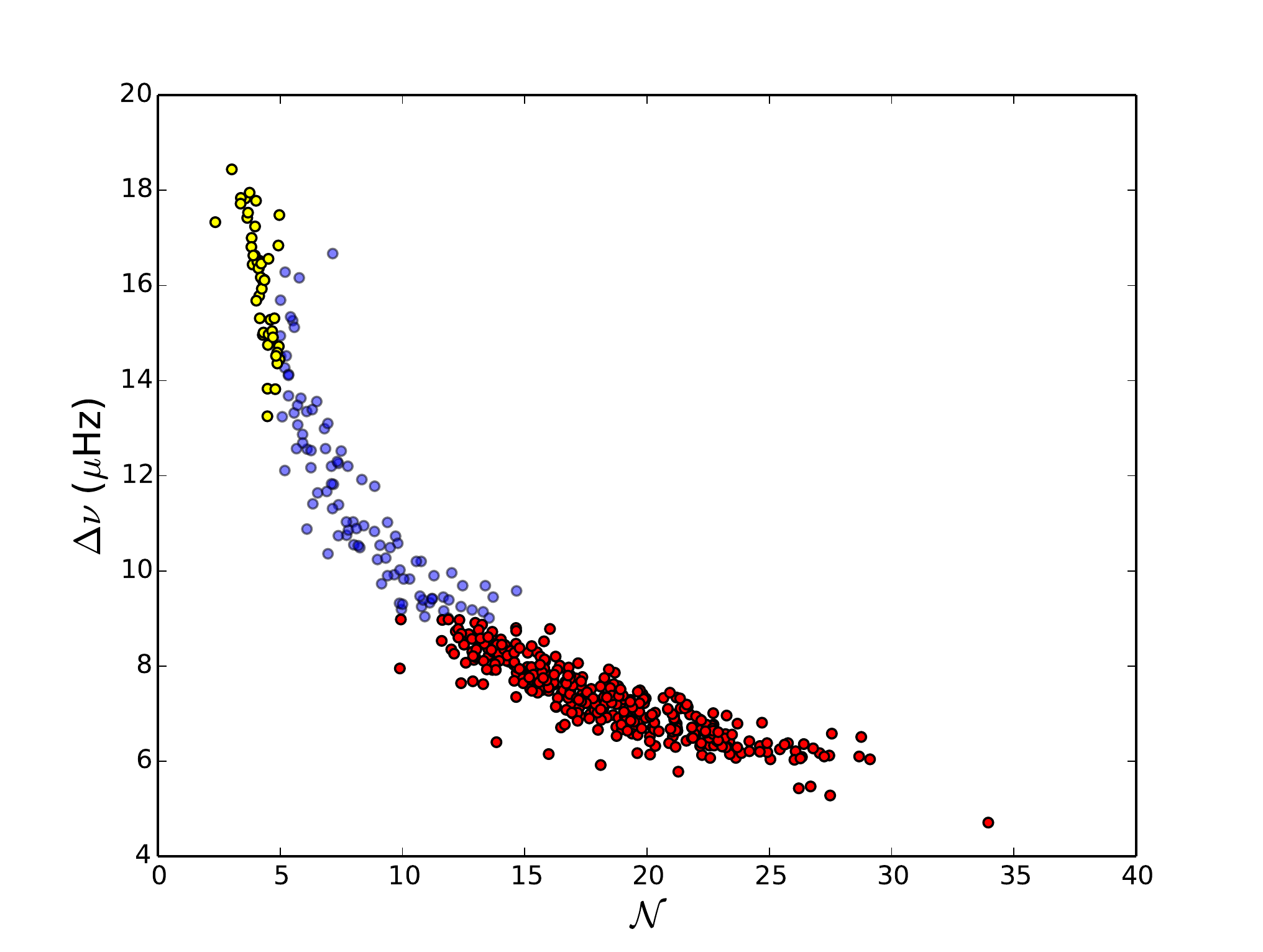}}
\caption{Large separation as a function of the mixed-mode density for red giant branch stars where our method failed to propose a satisfactorily identification of the rotational components. Stars having $\N \leq 5$ are represented in yellow, stars having $\Dnu \leq 9$ $\mu$Hz are represented in red.}
\label{fig-Dnu-N}
\end{figure}

\section{Large-scale measurements of the red giant core rotation in the \textit{Kepler} sample}\label{results}

We selected red giant branch stars where \cite{Vrard} obtained measurements of $\DPi$. These measurements were used as input values for $\Delta\Pi\ind{1,test}$ in the correlation method. The method proposed a satisfactory identification of the rotational multiplet components for 1183 red giant branch stars, which represents a success rate of 69\%. We obtained mean core rotation measurements for 875 stars on the red giant branch (Fig.~\ref{fig-dnurot}), roughly increasing the size of the sample by a factor of 10 compared to \cite{Mosser_2012c}. The impossibility to fit the rotational components increases when $\Dnu$ decreases: 70 \% of the unsuccessful cases correspond to $\Dnu \leq 9$ $\mu$Hz (Fig.~\ref{fig-Dnu-N}). Low $\Dnu$ values correspond to evolved red giant branch stars. During the evolution along the red giant branch, the g-m mode inertia increases and mixed modes are less visible, making the identification of the rotational components more difficult \citep{Dupret, Grosjean}. The method also failed to propose a satisfactorily identification of the rotational components where the mixed-mode density is too low: 10 \% of the unsuccessful cases correspond to $\N \leq 5$ (Fig.~\ref{fig-Dnu-N}). These cases correspond to stars on the low red giant branch presenting only few $\ell = 1$ g-m modes, where it is hard to retrieve significant information on the rotational components.

\subsection{Monitoring the evolution of the core rotation}\label{N-justif}

\begin{figure}
\resizebox{\hsize}{!}{\includegraphics{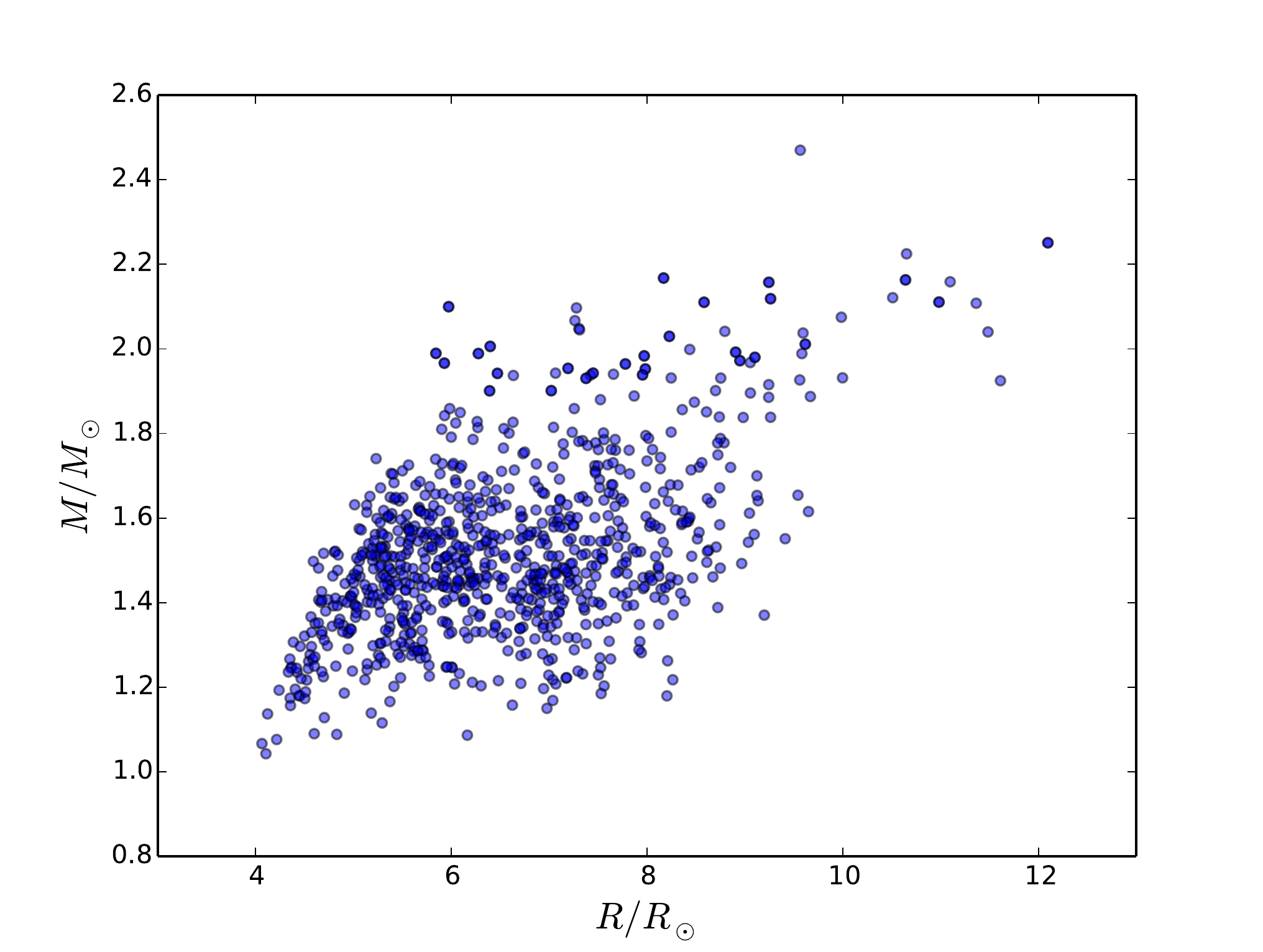}}
\caption{Mass as a function of the radius for red giant branch stars where the rotational multiplet components have been identified.}
\label{fig-M-R}
\end{figure}

\begin{figure}
\resizebox{\hsize}{!}{\includegraphics{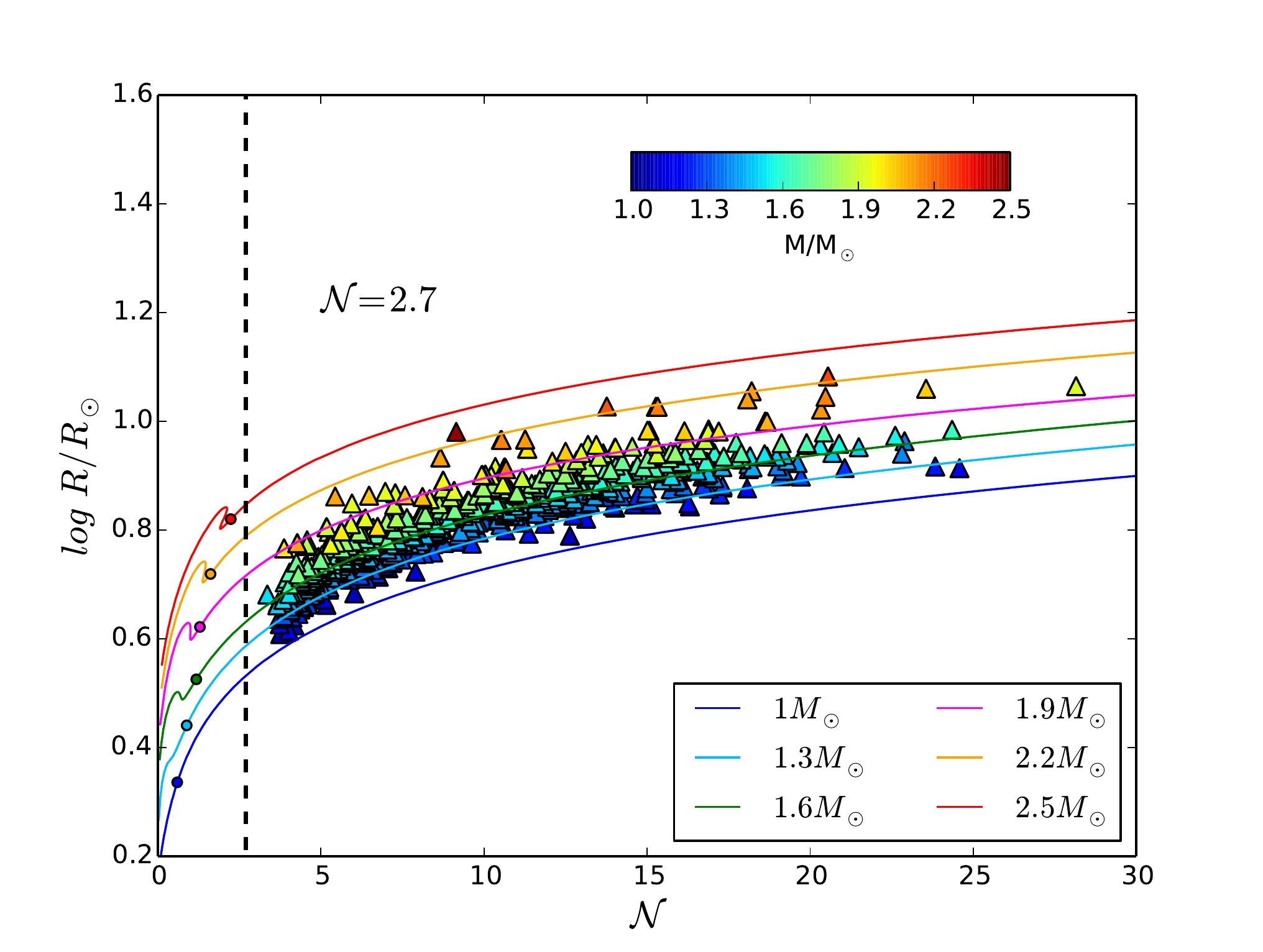}}
\caption{Evolution of the radius as a function of the mixed-mode density on the red giant branch. Colored triangles represent the measurements obtained in this study, the color coding the mass estimated from the asteroseismic global parameters. Evolutionary tracks are computed with MESA for different masses. The colored dots indicate the bottom of the red giant branch for the different masses. The vertical black dashed line indicates the lower observational limit for the mixed-mode density, $\N=2.7$.}
\label{fig-R}
\end{figure}

\begin{figure}
\resizebox{\hsize}{!}{\includegraphics{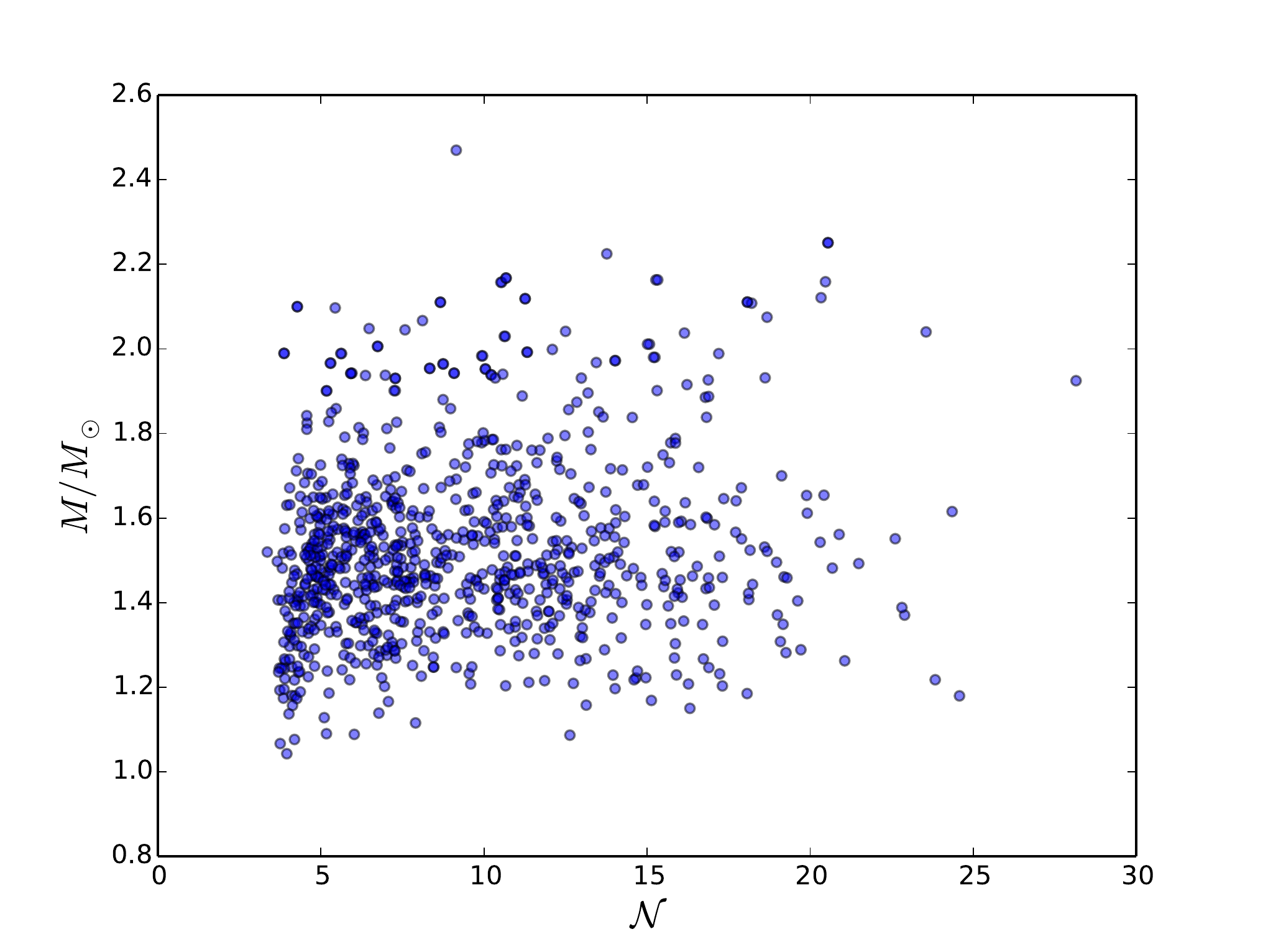}}
\caption{Same as Fig.~\ref{fig-M-R}, with now the mixed-mode density on the x-axis.}
\label{fig-M-N}
\end{figure}

\begin{figure}
\resizebox{\hsize}{!}{\includegraphics{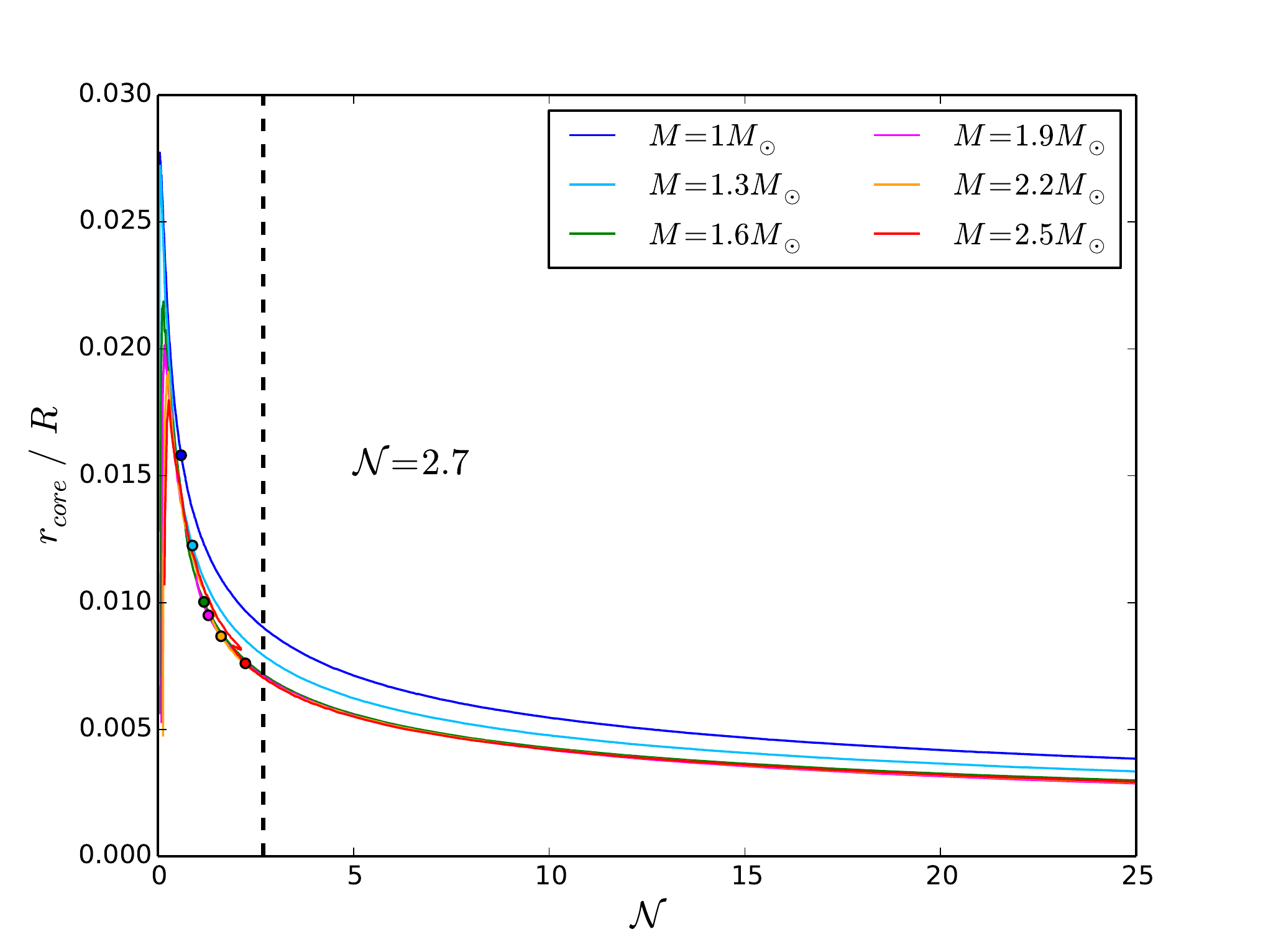}}
\caption{Evolution of the relative position of the core boundary, namely the radius of the core normalized to the total stellar radius, as a function of the mixed-mode density $\N$, for different stellar masses, computed with MESA. The color code for the evolutionary sequences is the same as in in Fig.~\ref{fig-R}. The vertical black dashed line has the same meaning as in Fig.~\ref{fig-R}. The tracks corresponding to $M \geq 1.6 \Msol$ are superimposed.}
\label{fig-R-fraction}
\end{figure}

The stellar masses and radii can be estimated from the global asteroseismic parameters $\Dnu$ and $\numax$ through the scaling relations \citep{Kjeldsen, Kallinger, Mosser_2013}
\begin{equation}\label{eqt-mass}
\frac{M}{\Msol} = \left(\frac{\numax}{\numaxsol}\right)^3 \left(\frac{\Dnu}{\Dnusol}\right)^{-4} \left(\frac{\Teff}{\Tsol}\right)^{3/2},
\end{equation}
\begin{equation}\label{eqt-radius}
\frac{R}{\Rsol} = \left(\frac{\numax}{\numaxsol}\right) \left(\frac{\Dnu}{\Dnusol}\right)^{-2} \left(\frac{\Teff}{\Tsol}\right)^{1/2},
\end{equation}
where $\numaxsol = 3050 \, \mu$Hz, $\Dnusol = 135.5 \, \mu$Hz and $\Tsol = 5777 \, K$ are the solar values chosen as references.
\newline
When available, we used the APOKASC effective temperatures obtained by spectroscopy \citep{Pinsonneault}, otherwise we used the scaling relation
\begin{equation}\label{eqt-Teff}
\Teff = 4800 \left(\frac{\numax}{40}\right)^{0.06},
\end{equation}
with $\numax$ in $\mu$Hz.
\newline
\newline
We observe a significant correlation between the stellar mass and radius in our sample (Fig.~\ref{fig-M-R}) with a Pearson correlation coefficient of 0.55, indicating that  the radius is not an appropriate parameter to monitor the evolution of the core rotation, as one can expect. Indeed, at fixed $\Dnu$, the higher the mass, the higher the expected radius on the red giant branch. In these conditions, the observed correlation between the stellar mass and radius is a bias induced by stellar evolution.
\newline
In order to illustrate this point, we computed evolutionary sequences for stellar models with $M = \{ 1.0, \, 1.3, \, 1.6, \, 1.9, \, 2.2, \, 2.5 \} \, \Msol$ with the stellar evolutionary code MESA \citep{Paxton_2011, Paxton_2013, Paxton_2015}. These models confirm that stars with higher masses enter the red giant branch with higher radii (Fig.~\ref{fig-R}). We further observe this trend in our sample when superimposing our data to the computed evolutionary tracks.
\newline
We stress that the mixed-mode density $\N$ is a possible proxy of stellar evolution instead of the radius, as models show that $\N$ increases when stars evolve along the red giant branch (Fig.~\ref{fig-R}). In fact, the computed evolutionary sequences indicate that while stars enter the red giant branch with a radius depending dramatically on the stellar mass, they show close $\N$ values between 0.6 and 2.6. We note that models confirm that all stars in our sample have already entered the red giant branch, as we cannot obtain seismic information for stars  below $\N = 2.7$ with
 \textit{Kepler} long-cadence data. We further observe an apparent absence of correlation between the stellar mass and the mixed-mode density in our sample (Fig.~\ref{fig-M-N}) with a Pearson correlation coefficient of 0.15, indicating that $\N$ is a less biased proxy of stellar evolution than the radius.
\newline
As shown by our models, the mixed-mode density remarkably monitors the fraction of the stellar radius occupied by the inert helium core along the red giant branch (Fig.~\ref{fig-R-fraction}). This is valid for the various stellar masses considered, as the relative difference in $r\ind{core}/R$ between models having different masses remains below 1 \%. We thus use the mixed-mode density as a proxy for stellar structure evolution on the red giant branch (Fig.~\ref{fig-dnurot}).


\subsection{Investigating the core slow-down rate as a function of the stellar mass}

\begin{figure*}
\centering
\includegraphics[width=12cm]{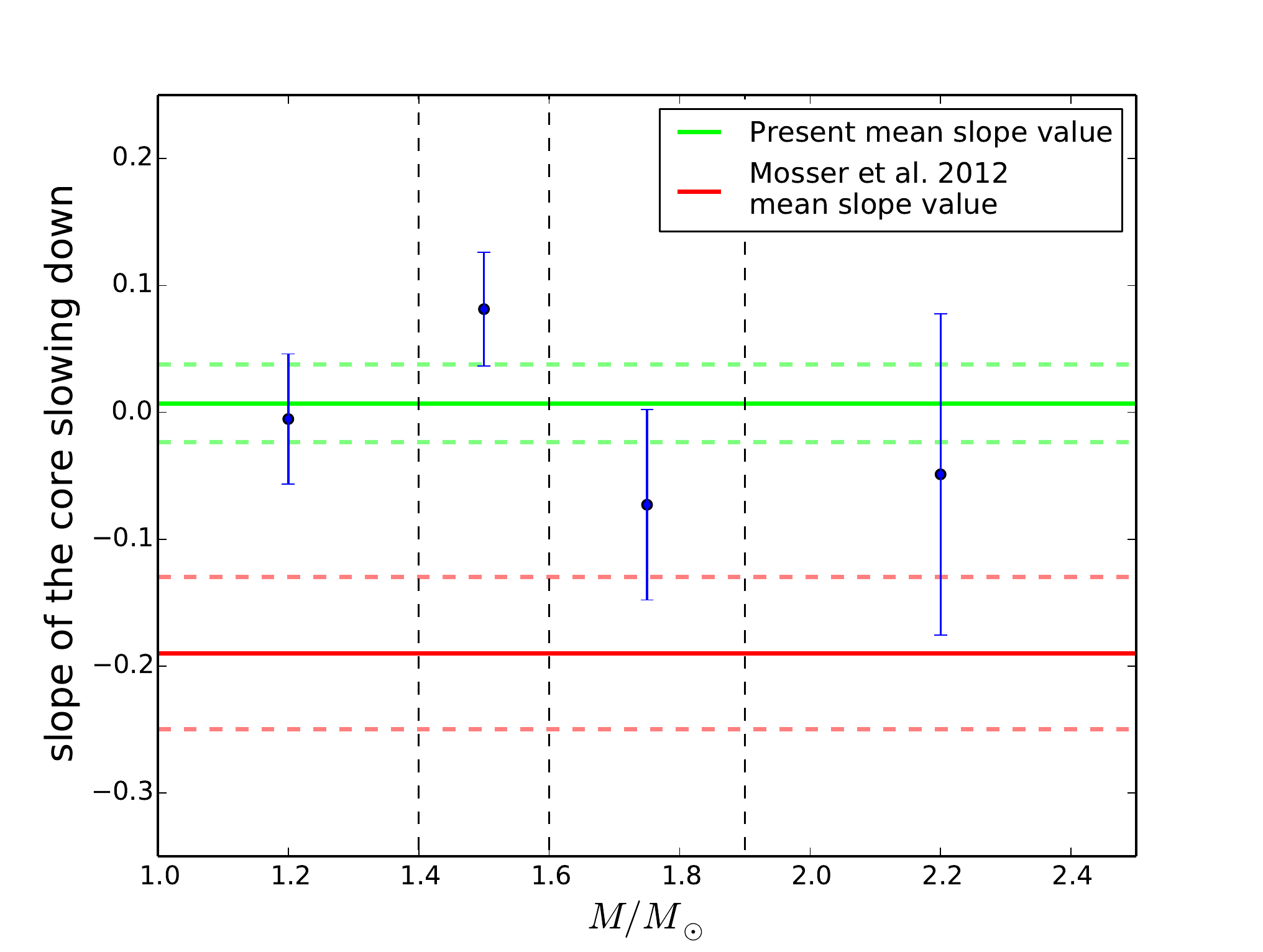}
\caption{Slopes of the core slow-down $a$ when considering the evolution of the core rotation as a function of the mixed-mode density $\N$ measured through Eq.~\ref{eqt-nmix} for different mass ranges. Our measurements and the associated error bars are represented in blue. Vertical black dashed lines mark the boundaries between the different mass ranges considered. The green and red solid and dashed lines indicate the value of the slow-down rate and the associated error bars measured in this study from a fit to all stars $\langle a \rangle$ and estimated from \cite{Mosser_2012c} measurements $a\ind{\N,2012}$, respectively.}
\label{fig-all-slopes}
\end{figure*}

On the one hand, \cite{Eggenberger_2017} explored the influence of the stellar mass on the efficiency of the angular momentum transport, but they only considered two stars with different masses. On the other hand, \cite{Mosser_2012c} measurements actually included a small number of stars on the red giant branch and the highest mass was around 1.7 $\Msol$, with only three high-mass stars.
We now have a much larger dataset covering a broad mass range, from 1 up to 2.5 $\Msol$, allowing us to investigate how the mean core rotational splitting and the slow-down rate of the core rotation depend on the stellar parameters. Considering different mass ranges, we measured each time the mean value of the core rotational splitting $\langle \dnurotcore \rangle$ and investigated a relationship of the type
\begin{equation}\label{eqt-mass-range-1}
\dnurotcore \propto \N^{a},
\end{equation}
with the $a$ values resulting from a non-linear least squares fit. The measured $\langle \dnurotcore \rangle$ and $a$ values are summarized in Table~\ref{table:1} as a function of the mass range (see also Fig.~\ref{fig-slopes} in Appendix \ref{A1}).
The results indicate that the mean core rotational splittings and core rotation slow-down rate are the same to the precision of our measurements for all stellar mass ranges considered in this study (Table~\ref{table:1} and Fig.~\ref{fig-all-slopes}). Moreover, the mean slow-down rate measured in this study is lower than what we measure when using \cite{Mosser_2012c} results (Table~\ref{table:2}).

\begin{table}
\caption{Measured slow-down rates and mean rotational splittings as a function of the mixed-mode density $\N$ for different mass ranges.}
\label{table:1}
\centering
\begin{tabular}{c c c}
\hline\hline
$M$ & $a$ & $\langle \dnurotcore \rangle$ (nHz)\\
\hline
$M \leq 1.4 \Msol$ & $-0.01 \pm 0.05$ & $331 \pm 127$ \\
$1.4 < M \leq 1.6 \Msol$ & $0.08 \pm 0.04$ & $355 \pm 140$ \\
$1.6 < M \leq 1.9 \Msol$ & $-0.07 \pm 0.07$ & $359 \pm 164$ \\
$M > 1.9 \Msol$ & $-0.05 \pm 0.13$ & $329 \pm 170$ \\
\hline
\end{tabular}
\end{table}

\section{Discussion}\label{discussion}

\begin{figure}[t]
\resizebox{\hsize}{!}{\includegraphics{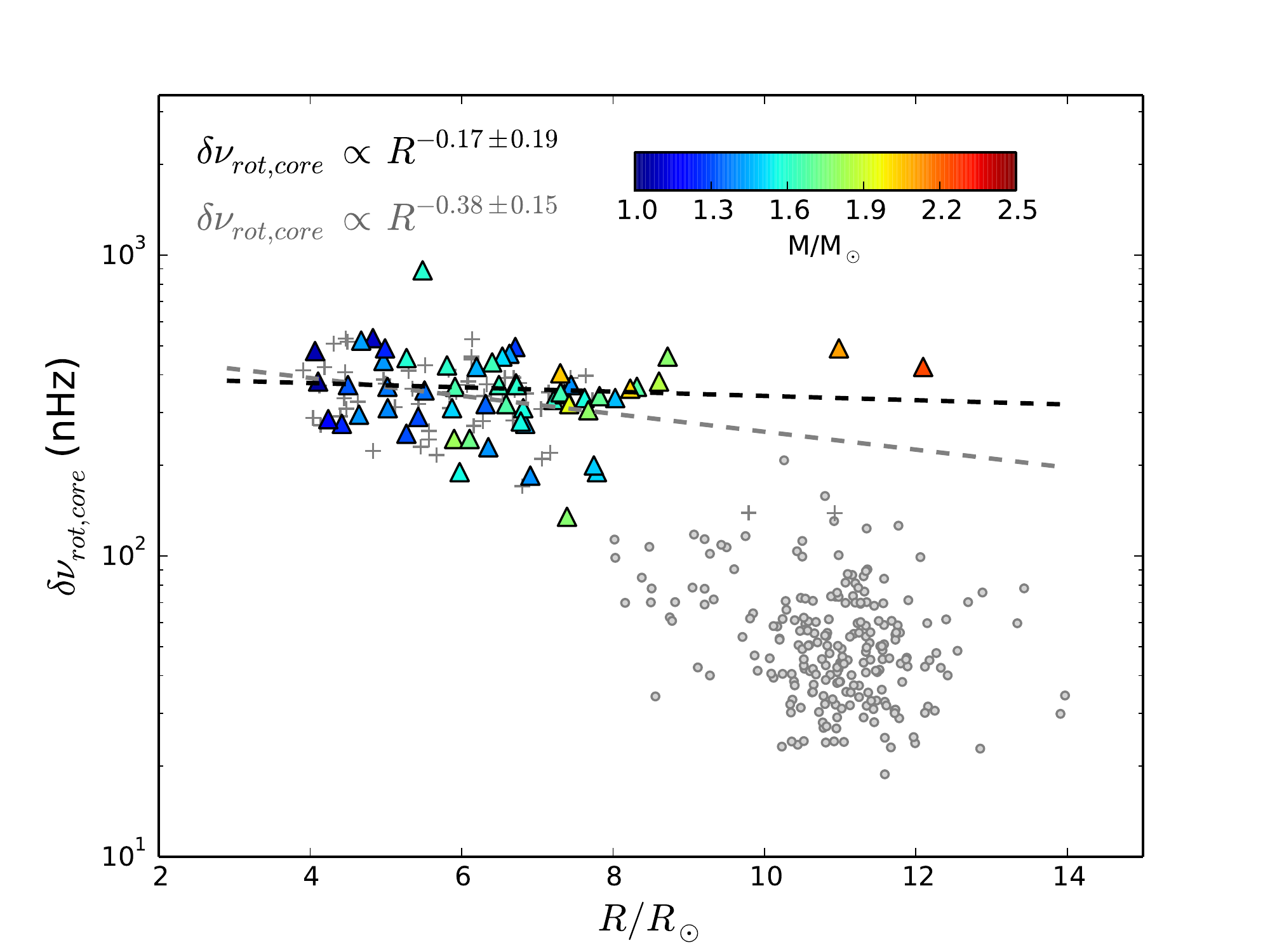}}
\caption{Mean core rotational splitting as a function of the radius estimated from the global asteroseismic parameters for the red giant branch stars studied by \cite{Mosser_2012c}. The color code of the colored triangles is the same as in Fig.~\ref{fig-dnurot}. \cite{Mosser_2012c} measurements on the red giant branch and on the clump are represented by grey crosses and dots, respectively. The linear fit resulting from the present measurements is plotted in black while the fit resulting from \cite{Mosser_2012c} measurements is represented in grey.}
\label{fig-R-2012}
\end{figure}

\begin{figure}[t]
\resizebox{\hsize}{!}{\includegraphics{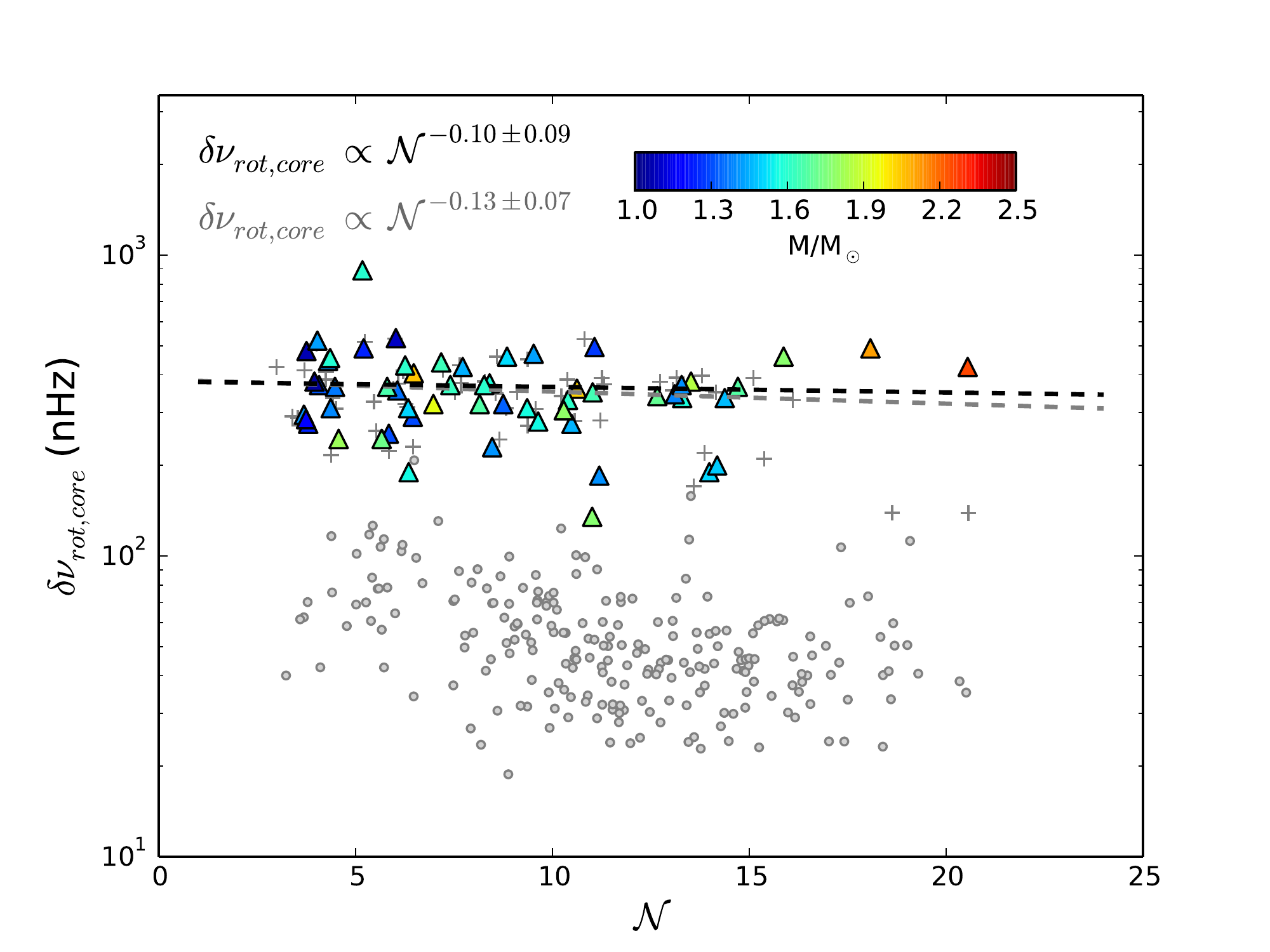}}
\caption{Same as Fig.~\ref{fig-R-2012}, this time representing the mean core rotational splitting as a function of the mixed-mode density for the red giant branch stars studied by \cite{Mosser_2012c}.}
\label{fig-N}
\end{figure}

We explore here the origin of the discrepancy found between the mean slow-down rate of the core rotation measured in this work and the mean slope measured when using \cite{Mosser_2012c} results. We selected the 57 red giant branch stars studied by \cite{Mosser_2012c} for which we obtained mean core rotation measurements with the method developed in this study and considered either the radius or the mixed-mode density as a proxy of stellar evolution.
\newline
\newline
We first measured the slow-down rates obtained with our measurements and those of \cite{Mosser_2012c} as a function of the radius (Fig.~\ref{fig-R-2012}). The measured slopes strongly differ from each other, the slow-down rate we obtained with our measurements being lower (Table~\ref{table:3}). 
The significant differences between these two sets of measurements rise from the two stars with a radius larger than 9.5 $\Rsol$, for which \cite{Mosser_2012c} underestimated the mean core rotational splittings. We checked that we recover slopes that are in agreement when excluding these stars from the two datasets.
\newline
\begin{table}[t]
\caption{Comparison of the measured slow-down rates as a function of the mixed-mode density $\N$ from the RGB stars in \cite{Mosser_2012c} sample ($a\ind{\N, 2012}$) and the measurements obtained in this study ($\langle a \rangle$), respectively.}
\label{table:2}
\centering
\begin{tabular}{c c}
\hline\hline
$a\ind{\N, 2012}$ & $\langle a \rangle$ \\
\hline
$-0.19 \pm 0.06$ & $0.01 \pm 0.03$ \\
\hline
\end{tabular}
\end{table}
\newline
We also made the same comparison using this time the mixed-mode density $\N$ as a proxy of stellar evolution (Fig.~\ref{fig-N}). We considered our measurements and those of \cite{Mosser_2012c} for the same set of stars and found slopes in agreement (Table~\ref{table:3}). The smaller discrepancy comes from the redistribution of the repartition of $\dnurotcore$ measurements induced by $\N$, which probes the stellar evolutionary stage, compared to what is observed when using the radius.
These slopes are in agreement but are significantly larger than what we derive from a sample ten times larger. The discrepancy thus comes from a sample effect. Hence, the sample of \cite{Mosser_2012c} is somehow biased. This is not surprising because their method was limited to simple cases, i.e. to small splittings. This confusion limit is more likely reached when $\Dnu$ decreases, which corresponds in average to more evolved stars. Thus, in the approach followed by \cite{Mosser_2012c}, it is easier to measure large splittings at low $\N$ since they are working with frequency instead of stretched periods. On the contrary, when $\N$ increases it is harder to measure large splittings because the rotational multiplet components are entangled. In these conditions, stars showing a non-ambiguous rotational signature are more likely to present low splittings. This deficit in large splittings at large $\N$ tends to result in a negative slope, suggesting a slow-down of the core rotation.
Working with stretched periods allows us to measure reliable splittings well beyond the confusion limit. Thus, our measurements, encompassing a much larger sample, allow to refine the diagnosis of \cite{Mosser_2012c} on the evolution of the core rotation along the red giant branch and to reveal that this evolution is independent of the stellar mass. They confirm low core-rotation rates on the low red giant branch, and indicate that the core rotation seems to remain constant on the part of the red giant branch covered by our sample, instead of sligthly slowing down. This reinforces the need to find physical mechanisms allowing to counterbalance the core contraction along the red giant branch, which should lead to an acceleration of the core rotation in the absence of angular momentum transport. Furthermore, models still need to reconcile with observations as they predict core rotation rates at least ten times higher than measured.
\newline
\begin{table}
\caption{Comparison of the measured slow-down rates as a function of different variables from the red giant branch stars in \cite{Mosser_2012c} sample, considering \cite{Mosser_2012c} measurements ($a\ind{2012, now}$) and the measurements obtained in this study ($a\ind{us}$).}
\label{table:3}
\centering
\begin{tabular}{c c c}
\hline\hline
Variable & $a\ind{2012, now}$ & $a\ind{us}$ \\
\hline
$R/\Rsol$ & $-0.38 \pm 0.15$ & $-0.17 \pm 0.19$ \\
$\N$ & $-0.13 \pm 0.07$ & $-0.10 \pm 0.09$ \\
\hline
\end{tabular}
\end{table}
\newline
Our results thus suggest a similar evolution of the core rotation for various masses on the ref giant branch. This constitutes a striking feature and hopefully a fruitful indication in the search for the physical process at work. It is not in contradiction with the conclusion made by \cite{Eggenberger_2017} that the efficiency of the angular momentum transport would increase with stellar mass, since a 2.5 $\Msol$ star evolves 100 times faster than a 1 $\Msol$ star on the red giant branch (Table~\ref{table:4}). However, things are not so simple as the evolutionary timescale is not the only parameter at stake. More modeling work will be necessary to go from these measurements to quantitative estimates in terms of angular momentum transport and to bring tighter constraints on the mechanisms at work. This goes beyond the scope of the present work.
\newline
\newline
Furthermore, we must keep in mind that the measured $\dnurotcore$ are the rotational splittings of the most g-dominated dipole modes, which does not directly scale with the core rotation rate.
In practice, $\bcore / \benv \gg 1$ only on the high red giant branch accessible with asteroseismology. On the lower giant branch, the envelope contribution to the splittings of g-m modes is not negligible. We thus need to introduce a correction factor $\eta$ in  Eq.~\ref{omega} to derive an accurate estimation of the mean core rotation rate \citep{Mosser_2012c}, as
\begin{equation}\label{omega-corr}
\omegacore = 4 \, \pi \, \eta \, \dnurotcore,
\end{equation}
with
\begin{equation}\label{eta}
\eta = 1 + \frac{0.65}{\N}.
\end{equation}
However, the relative deviation between $\omegacore$ values derived with and without the $\eta$ correction factor remains limited to around 15 \% in absolute value. In these conditions, $\dnurotcore$ can be used as a proxy of the red giant mean core rotation and provides clues on its evolution during the red giant branch stage.
Moreover, the $\eta$ correction is only a proxy for the correction of the envelope contribution. A complete interpretation in terms of the evolution of the red giant core rotation from the measured splittings requires the computation of rotational kernels, what is let to further studies.
\newline
\newline
We must also keep in mind that asteroseismology allows us to probe stellar interiors for only a fraction of the time spent by stars on the red giant branch ($t\ind{obs} / t\ind{RGB}$ in Table~\ref{table:4}). Thus, existing seismic measurements cannot bring constraints on the evolution of the core rotation on the high red giant branch, and we have to rely on modeling in this evolutionary stage. We know that a significant braking must occur in this evolutionary stage, unfortunately out of reach of seismic measurements. Indeed, compared to what we measure for red giant branch stars, the mean rotational splitting on the red clump is divided by a factor 8 for $M \leq 1.2 \Msol$ stars and a factor 4 for $M> 1.9 \Msol$ stars, as seen in Fig.~\ref{fig-dnurot}.
\newline
\begin{table}
\caption{Evolutionary timescales given by MESA for different stellar masses, either on the whole red giant branch ($t\ind{RGB}$) and accessible through asteroseismology for $3 < \N < 30$ ($t\ind{obs}$).}
\label{table:4}
\centering
\begin{tabular}{c c c c}
\hline\hline
$M/\Msol$ & $t\ind{RGB}$ (Myr) & $t\ind{obs}$ (Myr) & $t\ind{obs}/t\ind{RGB} (\%)$ \\
\hline
$1.0$ & $640$ & $200$ & $31$ \\
$1.3$ & $420$ & $160$ & $38$ \\
$1.6$ & $280$ & $120$ & $43$ \\
$1.9$ & $110$ & $50$ & $46$ \\
$2.2$ & $44$ & $13$ & $30$ \\
$2.5$ & $8$ & $6$ & $75$ \\
\hline
\end{tabular}
\end{table}
\newline
Finally, the fact that $\N$ is a good proxy of stellar evolution on the red giant branch does not stand anymore on the clump. Clump and red giant branch stars have mixed-mode densities located in the same range of values, which seems to indicate that $\N$ decreases from the tip of the red giant branch (Fig.~\ref{fig-dnurot}). This is confirmed by models computed with MESA which indicate that $\N$ strongly decreases when helium burning is firmly established in the core.
Moreover, data emphasize that the dispersion of the $\N$ values met in the clump  rises from a mass effect and does not trace the stellar evolution in the clump (Fig.~\ref{fig-dnurot}). Higher mass stars have lower $\N$ values in average. This trend is confirmed by models that predict lower $\N$ values on the clump when the stellar mass increases.

\section{Conclusion\label{conclusion}}

Mixed-modes and splittings can now be disentangled in a simple and almost automated way through the use of stretched period spectra. This opens the era of large-scale measurements of the core rotation of red giants, necessary to prepare the analysis of Plato data representing hundreds of thousands of potential red giants. We developed a method allowing an automatic identification of the dipole gravity-dominated mixed-modes split by rotation, providing us with a measurement of the mean core rotational splitting $\dnurotcore$ for red giant branch stars, even when mixed-modes and splittings are entangled. As rotational components having different azimuthal orders are now well identified in the full power spectrum, we can now have access to large rotational splitting values in a simple way.
\newline
\newline
As evolutionary sequences calculated with MESA for various stellar masses emphasized that the radius, used by \cite{Mosser_2012c}, is not a good proxy of stellar evolution, we used instead the mixed-mode density $\N$.
\newline
\newline
We obtained mean core rotation measurements for 875 red giant branch stars, covering a broad mass range from 1 to 2.5 $\Msol$. This led to the possibility of unveiling the role of the stellar mass in the core rotation slow-down rate. These measurements, obtained using a more elaborated method applied to a much larger sample, allowed to extend the results of \cite{Mosser_2012c} through a more precise characterization of the evolution of the mean core rotation on the red giant branch, and to unveil how this evolution depends on the stellar mass. Our results are not in contradiction with the conclusions of \cite{Mosser_2012c}, as they confirm low core rotation rates on the low red giant branch and indicate that the core rotation is almost constant instead of sligthly slowing down on the part of the red giant branch to which we have access.
It also appeared that this rotation is independent of the mass and remains constant over the red giant branch ecompassed by our observational set. This conclusion differs from previous results and is due to the combination of the use of a larger and less biased data set. As stars in our sample evolve on the red giant branch with very different timescales, this result implies that the mechanisms transporting angular momentum should have different efficiencies for different stellar masses. Quantifying the efficiency of the angular momentum transport requires a deeper use of models. This is beyond the scope of this paper and let to further studies.

\begin{acknowledgements}
We acknowledge financial support from the Programme National de Physique Stellaire (CNRS/INSU). We thank Sébastien Deheuvels for fruitful discussions, and the entire \textit{Kepler} team, whose efforts made these results possible.
\end{acknowledgements}

\bibliographystyle{./bibtex/aa} 
\bibliography{./bibtex/biblio}

\begin{appendix}
\section{Core slow-down rate as a function of the stellar mass}\label{A1}

\begin{figure*}
\includegraphics[width=9cm]{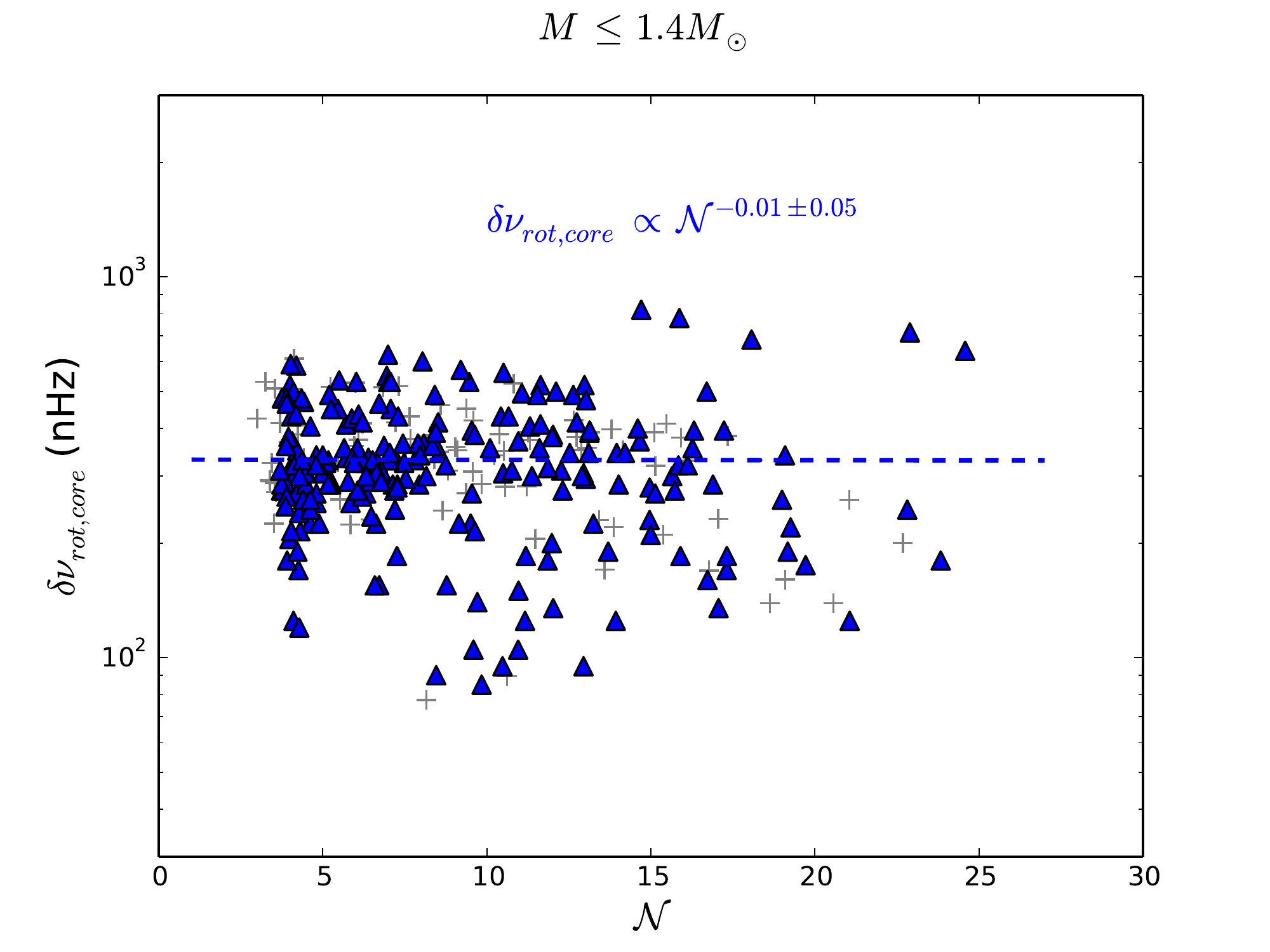}
\includegraphics[width=9cm]{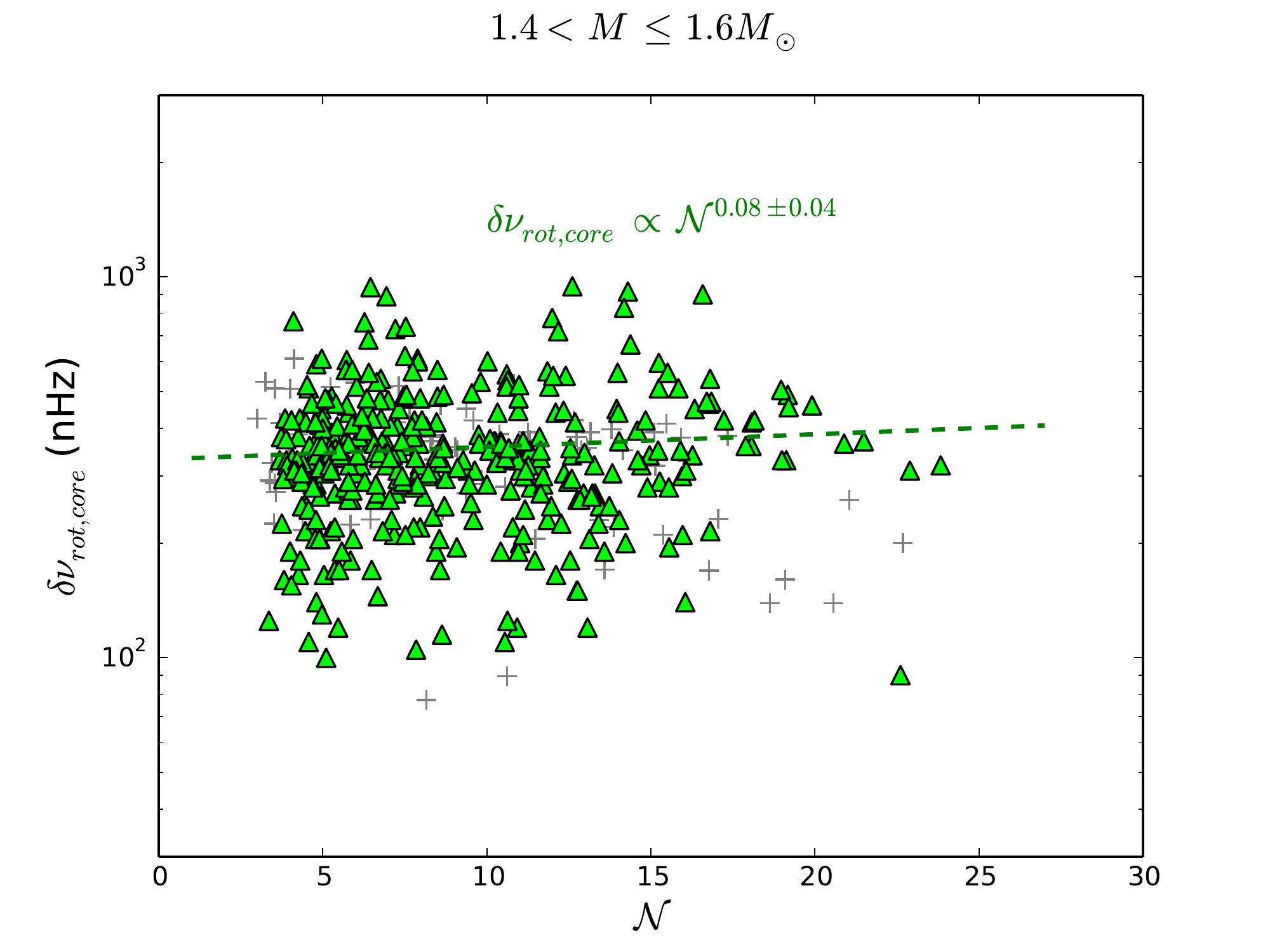}\\
\includegraphics[width=9cm]{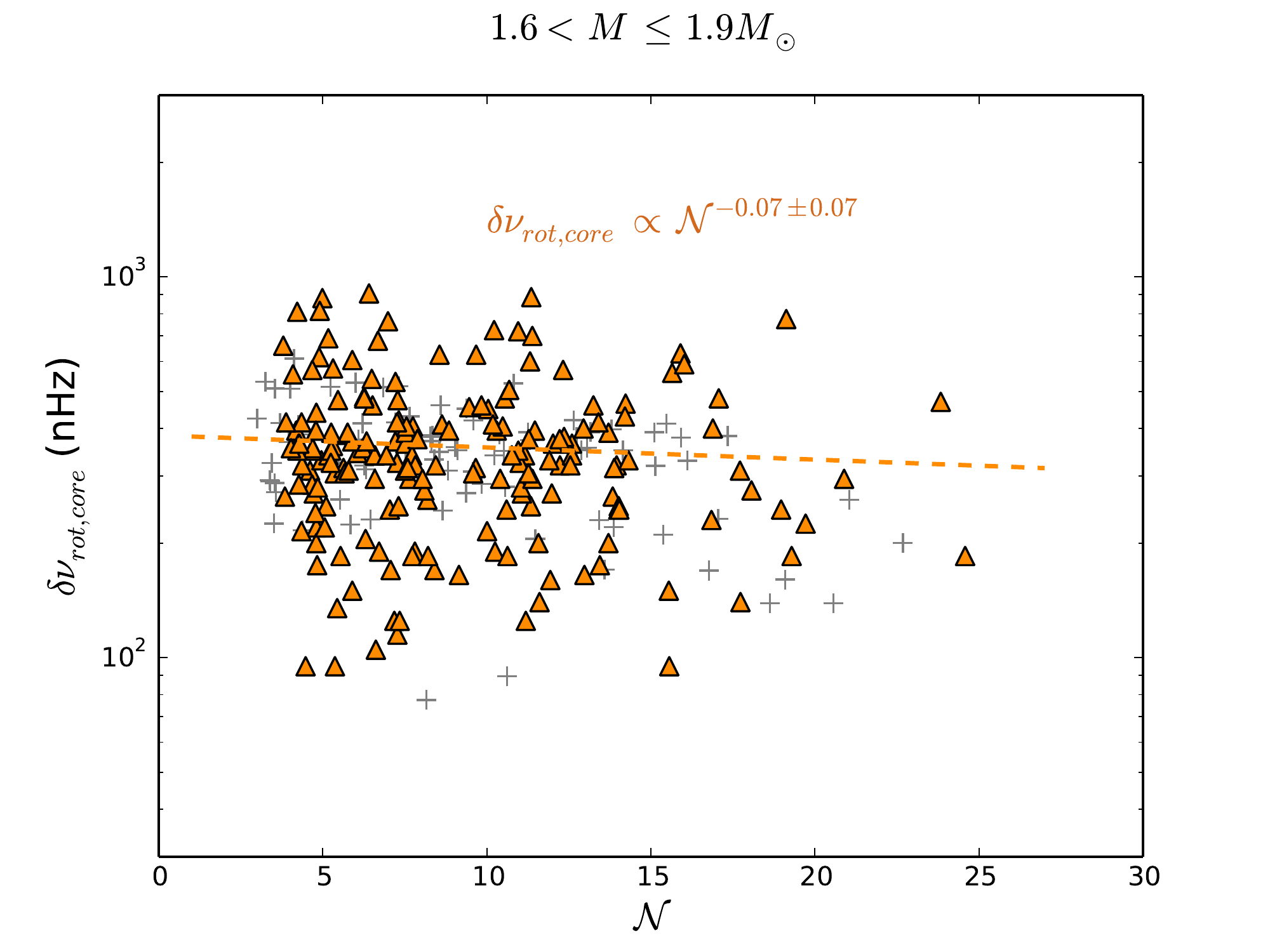}
\includegraphics[width=9cm]{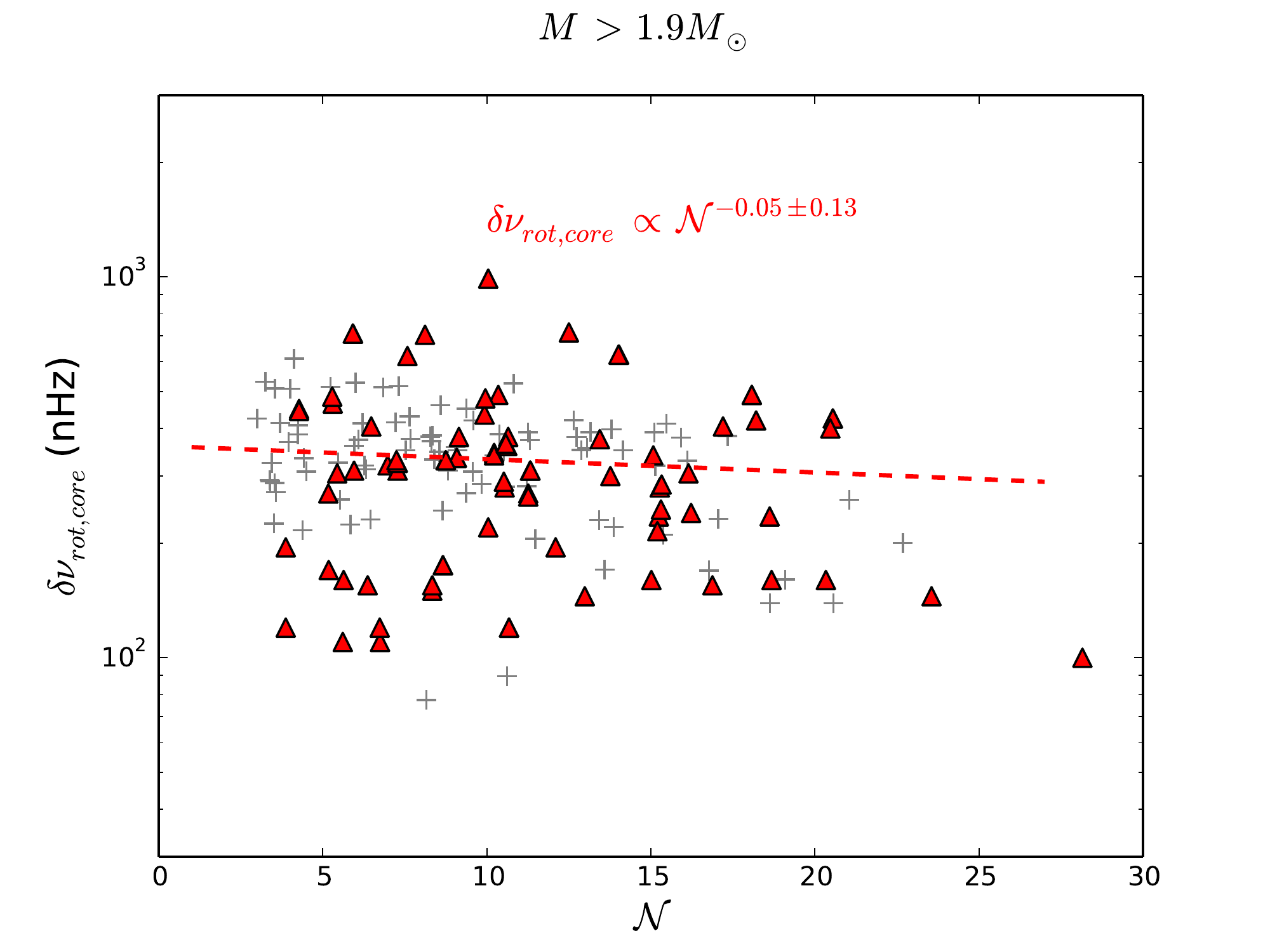}
\caption{Mean core rotational splitting as a function of the mixed-mode density for different mass ranges. Gray crosses represent the measurements of \cite{Mosser_2012c} while colored triangles represent the measurements obtained in this study. The colored lines represent the corresponding fits of the core slow-down. \textit{Upper left:} M $\leq$ 1.4 $\Msol$. \textit{Upper right:} 1.4 < M $\leq$ 1.6 $\Msol$. \textit{Lower left:} 1.6 < M $\leq$ 1.9 $\Msol$. \textit{Lower right:} M > 1.9 $\Msol$.}
\label{fig-slopes}
\end{figure*}

We have investigated the slowing-down of the core rotation as a function of the stellar mass, derived from its seismic proxy \citep{Kallinger}. The mass ranges were chosen in order to ensure a sufficiently large number of stars in each mass interval. As explained in the main text (Section \ref{N-justif}), the mixed-mode density is used as an unbiased marker of stellar evolution on the red giant branch. Results are shown in Fig.~\ref{fig-slopes}.

\end{appendix}

\end{document}